\documentclass[universe,article,accept,pdftex,moreauthors]{Definitions/mdpi} 

\firstpage{1} 
\pubvolume{1}
\issuenum{1}
\articlenumber{0}
\pubyear{2025}
\copyrightyear{2025}
\externaleditor{Firstname Lastname} 
\datereceived{6 November 2025} 
\daterevised{18 November 2025} 
\dateaccepted{24 November 2025} 
\datepublished{ } 

\hreflink{\textls[-45]{https://doi.org/}} 

\usepackage{wasysym}

\Title{Building a Radio AGN Sample from Cosmic \mbox{Morning---The }Radio High-Redshift Quasar Catalog (RHzQCat): I. Catalog from SDSS Quasars and Radio Surveys at $z \ge 3$}

\TitleCitation{Building a Radio AGN Sample from Cosmic Morning---The Radio High-Redshift Quasar Catalog (RHzQCat): I. Catalog from SDSS Quasars and Radio Surveys at $z \ge 3$}

\Author{Yingkang Zhang $^{1,2,}$*\orcidA{}, Ruqiu Lin $^{1,3}$\orcidD{}, Krisztina Perger $^{4,5}$\orcidB{}, S\'andor Frey $^{4,5,6}$\orcidC{}, Tao An $^{1,2}$\orcidE{}, Xiang Ji $^{1}$\orcidF{}, Qiqi Wu $^{1}$\orcidG{} \mbox{and Shilong Liao $^{1,7}$}\orcidH{}}

\AuthorNames{Yingkang Zhang, Ruqiu Lin, Krisztina Perger, S\'andor Frey, Tao An, Xiang Ji, Qiqi Wu, and Shilong Liao}

\AuthorCitation{Zhang, Y.; Lin, R.; Perger, K.; Frey, S.; An, T.; Ji, X.; Wu, Q.; Liao, S.}

\address{%
$^{1}$ \quad Shanghai Astronomical Observatory, Chinese Academy of Sciences, Nandan Road 80, \mbox{Shanghai 200030, China;} ruqiulin@umass.edu (R.L.); antao@shao.ac.cn (A.T.); jixiang@shao.ac.cn (X.J.); \mbox{wuqiqi@shao.ac.cn (Q.W.);} shilongliao@shao.ac.cn (S.L.)
\\
$^{2}$ \quad State Key Laboratory of Radio Astronomy and Technology, A20 Datun Road, Chaoyang District, Beijing, P. R. China\\
$^{3}$ \quad Department of Astronomy, University of Massachusetts, Amherst, MA 01003, USA\\
$^{4}$ \quad Konkoly Observatory, HUN-REN Research Centre for Astronomy and Earth Sciences, Konkoly Thege Mikl\'{o}s \'{u}t 15--17, 1121 Budapest, Hungary; perger.krisztina@csfk.org (K.P.); frey.sandor@csfk.org (S.F.) \\
$^{5}$ \quad CSFK, MTA Centre of Excellence, Konkoly Thege Mikl\'{o}s \'{u}t 15--17, 1121 Budapest, Hungary\\
$^{6}$ \quad Institute of Physics and Astronomy, ELTE E\"otv\"os Lor\'and University, P\'azm\'any P\'eter s\'et\'any 1/A, \mbox{1117 Budapest, Hungary}\\
$^{7}$ \quad School of Astronomy and Space Sciences, University of Chinese Academy of Sciences, No. 19A Yuquan Road, Beijing 100049, China\\
}

\corres{Correspondence: ykzhang@shao.ac.cn (Y.Z.)}

\abstract{
Radio-loud high-redshift quasars (RHRQs) provide crucial insights into the evolution of relativistic jets and their connection to the growth of supermassive black holes.
Beyond the extensively studied population at $z \ge 5$, the cosmic morning epoch ($3 \lesssim z \lesssim 5$) marks the peak of active galactic nucleus (AGN) activity and black hole accretion, yet remains relatively unexplored.
In this work, we compiled the radio high-redshift quasar catalog (RHzQCat) by cross-matching the SDSS DR16Q catalog with four major radio surveys---FIRST,NVSS, RACS, and GLEAM.
Our tier-based cross-matching framework and visual validation ensured reliable source identification across surveys with diverse beam sizes.
The catalog included 1629 reliable and 315 candidate RHRQs, with radio luminosities uniformly spanning $10^{25.5}$ -- $10^{29.3}$ W Hz$^{-1}$.
About 95\% of the confirmed sources exhibited compact morphologies, consistent with Doppler-boosted or young AGN populations at high redshifts.
Our catalog increases the number of known RHRQs at $z\ge3$ by an order of magnitude, representing the largest and most homogeneous catalog of radio quasars at cosmic morning, filling the observational gap between the early ($z>6$) and local Universe.
It provides a robust reference for future statistical studies of jet evolution, AGN feedback, and cosmic magnetism with next-generation facilities such as the Square \mbox{Kilometer Array (SKA).}
}


\begin{document}

\section{Introduction} \label{sec1:intro}

Active galactic nuclei (AGNs) are among the most energetic persistent extragalactic objects, with their immense power primarily originating from the accretion process of their central supermassive black holes (SMBHs).
Due to their distinctive and prominent emission across the entire electromagnetic spectrum, AGNs are considered ideal probes of the Universe. Particularly, the investigation of high-redshift quasars (HRQs) is extremely important to understand the early stages of galaxy formation and the evolution of SMBHs, as these objects most probably represent the first generations of massive accreting systems. They can produce powerful jets that interact with the early intergalactic medium (IGM), providing critical insights into the ionization history of IGM, the build-up of massive galaxies, and the feedback processes driven by AGNs \cite{2014MNRAS.442L..81F,2016ARA&A..54..313M,2019ARA&A..57..467B,2020NatCo..11..143A,2024MNRAS.528.5692K}.

The first samples of HRQs were discovered in the early 2000s, and their spectral characteristics offered strong proof of cosmic re-ionization \cite{2006AJ....132..117F}.
Spectroscopic redshifts at $z\ge5$ are mainly from the Sloan Digital Sky Survey (SDSS) \cite{2002AJ....123..485S} and the PanSTARRS1 (Panoramic Survey Telescope and Rapid Response System)
 (PS1) \cite{2002SPIE.4836..154K,2010SPIE.7733E..0EK} survey. At higher redshifts ($z\ge6$), values measured by near-infrared (NIR) instruments and surveys become dominant \cite{2020MNRAS.494..789R,2025MNRAS.540.1308O}.
To date, the population of HRQs (e.g., $z>5$) has steadily increased to over 600 \cite{2010AJ....139..906W,2017AJ....153..184Y,2023ApJS..269...27Y,2020MNRAS.494..789R,2025MNRAS.540.1308O}. 

As a radio subclass of the HRQ population, radio-loud HRQs (RHRQs) represent the most intriguing source sample for several reasons:
(1) their radio emission is not affected by dust extinction and can penetrate dense interstellar and intergalactic environments; \mbox{(2) the} central engines of radio-loud quasars represent the most active and energetic accreting objects in the Universe, and their evolution across cosmic time is crucial for revealing the growth histories of SMBHs and the evolution of the cosmic environment; and (3) radio jets are unique probes for the study of early SMBH environments and cosmological models \cite{2020NatCo..11..143A,2020A&A...643L..12S,2022ApJ...937...19Z,2025A&A...698A.158I,2025NatAs...9..293B,1993Natur.361..134K} as their radio structure can be investigated down to parsec (pc) or sub-pc scales thanks to the very long baseline interferometry (VLBI) technique. 
Moreover, due to the fact that RHRQs constitute only approximately $10\%$ \cite{2002AJ....124.2364I,2016ApJ...831..168K,2024MNRAS.528.5692K,2025ApJS..280...23A} of the quasar population, they are particularly rare and valuable, especially at high redshifts where the total number of known HRQs is smaller. 

Despite the intensive efforts devoted to searching for and studying quasars at $z > 5$, the population at $3 \lesssim z \lesssim 5$ has been relatively less explored, yet remains crucial for understanding AGN evolution.
This "cosmic morning" era coincides with the most active phase of SMBH growth, linking the early Universe to the local Universe \citep{2020ApJ...897..177P}.
During this period, flat-spectrum radio quasars (FSRQs) dominate the blazar population, and even stand out among all AGNs due to selection biases.
Studying blazars within the HRQ population at $z > 3$ offers a powerful means to constrain the evolution of relativistic jets and their coupling with the central engine or the black hole growth \cite{2014MNRAS.442L..81F,2017ApJ...836L...1T}.
Progress in this direction will require coordinated multi-frequency and high-resolution radio observations, along with statistically large samples, to trace jet properties and energetics across \mbox{cosmic time.}

In recent years, the identification of RHRQs has gained significant momentum, largely driven by the rapid advancements in both wide-field optical spectroscopic surveys and high-sensitivity radio surveys. Perger et al. \cite{2017FrASS...4....9P} (hereafter P17)
 built an updated RHRQ catalog for quasars at $z\ge4$, enabling follow-up studies uncovering the weak radio quasar population \cite{2024MNRAS.527.3436P} and VLBI analysis of selected individual objects \cite{2022ApJS..260...49K}.

Current studies of RHRQs still rely heavily on combining optical spectroscopic surveys with radio surveys \cite{2024A&A...689A.174C}.
The matching criteria adopted are generally simple, often based on the literature values or a fixed conservative positional difference.
While such approaches have yielded important insights---particularly for targeted analyses of small RHRQ subsets---they are inherently biased toward radio surveys with a high (arcsecond-scale) angular resolution, such as the Faint Images of the Radio Sky at Twenty Centimeters (FIRST) \citep{1995ApJ...450..559B}, 
the \mbox{Karl G.} Jansky Very Large Array Sky Survey (VLASS) \citep{2020PASP..132c5001L}, and
the Rapid ASKAP Continuum Surveys (the RACS series) \citep{2021PASA...38...58H,2024PASA...41....3D,2025PASA...42...38D},
and become increasingly inadequate when applied to radio sky surveys with poorer resolutions.

In particular, due to the simplicity and conservatism of traditional cross-matching strategies, a fraction of genuine optical--radio counterparts can be missed, while spurious associations may be introduced---particularly in the cases of radio surveys with large beam sizes such as 
the NRAO VLA Sky Survey (NVSS) \citep{1998AJ....115.1693C} and the GaLactic and Extragalactic All-sky MWA Survey (GLEAM) \citep{2017MNRAS.464.1146H}.
In such cases, both the source identification and the flux density measurements are prone to significant uncertainties, which can result in false matches and biased spectral properties.
These limitations highlight the necessity of adopting a more reliable, multi-tiered matching framework, especially for surveys characterized by a low angular resolution and large positional uncertainties.

In this paper, we constructed the radio high-z quasar catalog (RHzQCat) based on the SDSS catalog \citep{2020ApJS..250....8L} and four major radio surveys (FIRST, NVSS, RACS, and GLEAM), forming the largest sample of radio quasars at $z>3$ to date.
We applied a novel cross-matching framework that leverages the differing angular resolutions of the radio surveys to identify all plausible RHRQs while minimizing the risk of spurious associations within the SDSS footprint.
This work also provides a foundation for future searches in forthcoming wide-field radio surveys conducted with next-generation radio telescopes, such as the Square Kilometer Array (SKA) and the next-generation VLA (ngVLA), which are expected to deliver both higher-resolution and deeper observations, significantly increasing the number of known high-redshift radio quasars.

Section~\ref{sec2:method} introduces the comprehensive cross-matching procedure for compiling the catalog. Section~\ref{sec3:catalog} gives a detailed description of the final catalog. In Section~\ref{sec4:discussion}, we discuss the radio characteristics of the RHRQ sources in our catalog and compare our results with previous studies; scientific aspects are also discussed in this section. Section~\ref{sec5:conclusion} provides a summary of the paper.

\section{Methods} \label{sec2:method}
\subsection{Building the Catalog}
To find high-redshift radio quasars, we adopted the SDSS Data Release 16 quasar catalog \citep{2020ApJS..250....8L} (hereafter DR16Q)
 as our primary source catalog of optical quasars. This dataset provides high-confidence spectroscopic redshifts for over 750,000 quasars across the northern sky.

With our primary aim of maximizing matching completeness, we intended to include all possible radio emitters from the parent optical survey so that single matches and group matches can be clearly distinguished. Among the major SDSS classifications (QSO, GALAXY, STAR) \citep{2020ApJS..249....3A}, stars are the least likely to produce a detectable radio emission in current surveys \citep{2002AJ....124.2364I,2016A&ARv..24...13P}. Optical galaxies, however, can host radio-loud AGNs \citep{1992ARA&A..30..575C,2016A&ARv..24...13P}, so we also included galaxies from SDSS DR16 in our parent catalog.

Because galaxies are extremely numerous in the optical sky (hundreds of millions), downloading and matching the full sample would be computationally prohibitive. To alleviate this, we only included the ``clean'' galaxies from SDSS DR16 \citep{2020ApJS..249....3A} (parameterized by \texttt{clean = 1} and \texttt{class = `galaxy'})\endnote{The majority of SDSS galaxies have no spectroscopic identifiers but only photometric redshifts. To include this huge subclass as well, we first performed cross-matching without this subclass, constructed a matched sample, and then performed an additional cross-match restricted to these photometric galaxies located near the matched sample. The resulting set was concatenated back into the parent optical catalog for the final cross-matching.}.

For the radio counterparts, we used four major survey catalogs: FIRST, NVSS, GLEAM, and RACS-DR1 \citep{2021PASA...38...58H}. These radio surveys differ in their beam sizes and positional accuracy. To preserve this information, we did not merge the radio catalogs at this stage. Instead, the optical catalog was cross-matched separately against each radio survey, and the resulting high-$z$ quasars were flagged and then combined into a unified matched catalog. The full matching and processing procedure is described below.

\begin{enumerate}
    \item Generating optical catalog. We compiled the optical parent catalog by combining DR16Q and the galaxies from DR16. The selection of galaxies was carried out by setting \texttt{class = `galaxy'}
 and \texttt{clean = 1} 
 in DR16.
    
    \item Cross-matching. We performed pairwise cross-matching between DR16Q and each of the radio catalogs. Due to differing angular resolutions and survey configurations, the matching radius was adjusted accordingly for each survey. 
    The second column in Table~\ref{tab1:match_parms} lists the matching parameter, in units of arcseconds (arcsec, $^{\prime\prime}$), for each radio--optical catalog match.

\begin{table}[H] 
\caption{The matching parameters used in the script.
\label{tab1:match_parms}}
\begin{tabularx}{\textwidth}{m{3.5cm}<{\raggedright}m{3.5cm}<{\centering}m{6cm}<{\raggedright}}
\toprule
\textbf{Matched Surveys} & \textbf{Matching Radius \textsuperscript{1}}	& \textbf{Columns for Secondary Filter \textsuperscript{2}}\\
\midrule
DR16Q--FIRST		& 3     	& Maj\\
DR16Q--NVSS		& 25		& MajAxis\\
DR16Q--GLEAM		& 110		& awide, psfawide \\
DR16Q--RACS-DR1	& 15		& amaj0, amaj \\
\bottomrule
\end{tabularx}
\noindent{\footnotesize{\textsuperscript{1} The matching radius used for the cross-match in units of arcsec. \textsuperscript{2} The names of columns from the original radio survey catalogs that are used for additional filtering.}}
\end{table}

    \item Additional filtering. Following the initial matches, we applied an additional filtering criterion based on spatial association.
    Radio sources whose separation from the optical quasar position exceeded the full width at half-maximum (FWHM) of the source size (i.e., the best source major axis; see text and Table~\ref{tab1:match_parms} third column) were excluded to mitigate false matches arising from positional uncertainties.

    \item Flagging. We flagged the matched sources into three tiers: \textbf{Tier 1} includes quasars with unambiguous one-to-one matches; \textbf{Tier 2} consists of source groups where all sources have a redshift above 3, suggesting possible multi-source systems at high redshifts; and \textbf{Tier 3} includes the group of sources that have mixed redshifts, where only one source or part of the sources are HRQs. Sources within this tier are likely contaminated by foreground objects.

    \item Combination. After the matching and flagging steps above, for each radio survey, we obtained their high-$z$ radio sample. We then combined the four matched radio samples as one single table, with each entry annotated with its tier flag from \mbox{survey-specific matches.}

\end{enumerate}

For the choice of matching radius, we considered the convolving beam sizes or psf (point spread function) for setting our matching radius (Table~\ref{tab1:match_parms}). For FIRST, NVSS, and RACS-DR1, fixed beam sizes of $5.4^{\prime\prime}$, $45^{\prime\prime}$, and $25^{\prime\prime}$ were applied, respectively. For GLEAM, we took the maximal value of 210$^{\prime\prime}$ based on the given psf ranges \cite{2017MNRAS.464.1146H}.
Further filtering the matched sources is crucial for differentiating between individual matches and grouped sources. At this stage, we used variable values for the matching radii. 
When a radio source is clearly resolved, its deconvolved source size can better reflect the physical origin of the radio emission. Thus, we used the following methods to select the best source major axis. For FIRST, the \texttt{Maj} 
 column was used when it was not zero and was smaller than the psf ($5.4^{\prime\prime}$) \citep{1997ApJ...475..479W}. For NVSS, the \texttt{MajAxis} 
 was used when it was smaller than its typical psf of $45^{\prime\prime}$. For GLEAM, where the beam is large and the psf shows obvious variability, we chose the smaller value between \texttt{awide} 
 and \texttt{psfawide}. 
 For RACS-DR1, we used the deconvolved major axis \texttt{amaj0} 
 when it was available. Otherwise, the smaller of the fitted major axis \texttt{amaj} 
 and the typical $25^{\prime\prime}$ psf was used.

As shown in the flagging step, we adopted a tier-based flag to demonstrate the grouping status of each matched source. 
For each radio survey, the number of sources with different tier flags are listed in Table~\ref{tab2:statistics}
(78 sources with unreliable redshifts are excluded, see Section \ref{subsec:redshifts} for details).
The tier flag can serve as an indicator, showing the degree of ambiguity for the
optical--radio associations for each radio catalog.
As shown in the table, the small beam of FIRST ($5.4^{\prime\prime}$) yields few contaminants when cross-matching to optical quasars and galaxies.
In contrast, large-beam surveys such as GLEAM ($\approx$110$^{\prime\prime}$ matching radius) rarely yield single, unambiguous matches. These tier flags therefore play an important role in guiding subsequent inspections. For example, sources with \mbox{Tier 1} (e.g., single, compact matches) are often straightforward, while higher-tier cases (e.g., multiple associations or extended morphologies) indicate cases where additional \mbox{scrutiny is warranted.}

\begin{table}[H] 
\caption{The number of sources from different match tiers in the matched catalog.
\label{tab2:statistics}}
\footnotesize
\begin{tabularx}{\textwidth}{m{3.5cm}<{\centering}CCm{2cm}<{\centering}Cm{1cm}<{\centering}}
\toprule
\textbf{Sources with Match Tiers} & \textbf{From FIRST}	& \textbf{From NVSS} & \textbf{From GLEAM} & \textbf{Rrom RACS} & \textbf{All}\\
\midrule
Total matched sources \textsuperscript{1} & 1607 & 1614 & 370 & 955 & 2310  \\
Tier 1 sources & 1603 & 831 & 0 & 782 &  / \\
Tier 2 sources & 0    & 2    & 0   & 0 &  / \\
Tier 3 sources & 4    & 781   & 370 & 173 &  / \\
\bottomrule
\end{tabularx}
\noindent{\footnotesize{\textsuperscript{1} The number of sources refers to the SDSS objects. The count of radio-detected sources is likely lower, as a single radio source might correspond to multiple SDSS objects at the Tier 2 or Tier 3 levels.}} 

\end{table}

\vspace{-8pt}
During the cross-matching, we employed \texttt{STILTS} 
 \cite{2006ASPC..351..666T} and \texttt{TOPCAT} 
 \cite{2005ASPC..347...29T} software to script the detailed cross-matching process, which are both user-friendly and suitable for Unix systems. The related scripts are available in a Github repository\endnote{\url{https://github.com/ykzhang112233/highz-catalogue}, accessed on 1 June 2025\label{link1}
}. This offers a practical and consistent approach to cross-matching, facilitating its application for diverse scientific objectives.

\subsection{Verification and Classification}
After compiling the matched catalog, further verification of the sources was still needed.
Firstly, the redshifts from DR16Q are not always highly reliable, especially for the high-redshift subsample. These sources are weak, and emission lines are difficult to identify within the spectral coverage.
Thus, the redshifts might need to be modified.
Secondly, for the optical--radio source matching, the \texttt{Tier} 
 flag can help us indicate the quality of cross-identification to some extent, but this is often not enough.

As the beam size increases, more contaminating sources can be found within this angular separation. These contaminants are mostly foreground galaxies, which could cause the following effects to a radio source: (1) they are not radio emitters and will not contribute to the radio emission of the target source; (2) some of them are weak radio sources, and the physically unrelated objects that are located within the matching radius may contribute extra radio emissions and cause the flux density of the target source to be overestimated; and (3) some of them are strong radio sources that may dominate the radio flux density, and in this case, the radio source is not the target RHRQ, but the foreground object instead.

Thus, we performed the following steps to further verify the catalog: 
(1) the source redshifts were verified based on their quality parameters in DR16Q, as well as via visual inspection on some particular source spectra;
(2) after redshift verification, we made optical--radio overlay plots for the remaining candidate sources, with an angular scale of 4$^{\prime\prime}$ $\times$ 4$^{\prime\prime}$, centered on the target position, to visualize the matching correspondence for the target sources.
Details of the verification are given below.

\subsubsection{Redshift Verification}
\label{subsec:redshifts}
DR16Q provides a hierarchical reliability analysis of their redshifts, all of which are derived from optical spectroscopy. In DR16Q, redshifts from visual inspection (\texttt{VI}) 
 are the most favorable; when there are no \texttt{VI} 
 sources or the quality of \texttt{VI} 
 is not good enough (i.e., small \texttt{Z\_CONF} 
 values in the original catalog), the \texttt{PIPE} 
 sources are favored (see \cite{2020ApJS..250....8L} for the full selection criteria on ``best'' redshifts).
The \texttt{PIPE} 
 redshifts without warning flags (i.e., \texttt{ZWARNING} 
 $=$ 0\endnote{A binary indicator showing the quality of the pipeline redshifts; if not 0, issues will be reported, see
 \url{http://www.sdss.org/dr16/algorithms/bitmasks/\#ZWARNING}}) are confirmed as reliable through certain validation tests in DR16Q. However, due to the relatively limited depth and spectral range \mbox{(3600--10,400~$\text{\AA}$)} of the SDSS, significant spectral features (like Ly$_{\alpha}$) of HRQs are typically either not observable or shifted beyond the SDSS spectral coverage, particularly at $z>5$. 
In some studies employing SDSS data, sources at redshifts higher than 5 are frequently omitted when relying on pipeline-generated redshifts due to their significant uncertainty \citep{2020ApJS..250....8L,2024ApJS..275...19Y}.

Given the context above, the reliability of redshifts must be carefully evaluated, particularly for our RHRQ samples.
In the compiled catalog, the \texttt{sdss\_z} 
 column reflects the ``primary'' redshifts derived from the ``best'' approach \texttt{sdss\_source\_z} 
 from DR16Q. These redshifts are mainly \texttt{VI} 
 (from visual inspection), \texttt{PIPE} 
 (from pipeline), or visually inspected ones from a previous data release (e.g., \texttt{DRxQ\_xx}, 
 where \texttt{xx} 
 denotes the serial number of the previous SDSS quasar catalog release; see the DR16Q paper \cite{2020ApJS..250....8L} or the original catalog for more details).

Since our catalog contains a number of sources using \texttt{PIPE} 
 as their redshift sources, we performed visual inspection on some spectra of these sources, and arrived at the following conclusions:
(1) When the ``best'' redshift of a source is \texttt{PIPE} 
 and its \texttt{ZWARNING} 
 is not zero, the redshifts are not reliable. For these sources, the most likely issues are low signal-to-noise ratios or poorly fitted results.
(2) When the ``best'' redshift of a source is \texttt{PIPE} 
 and \texttt{ZWARNING = 0}, 
 significant emission lines can be seen, while the determination of redshifts from the pipeline is still ambiguous, mostly due to the misidentification of certain lines. During our visual inspection of some of these sources, we found that some redshifts are correct, some are misidentified, and some cannot be determined, requiring follow-up infrared spectral information.

\begin{figure}[H]
\includegraphics[width=6.2cm]{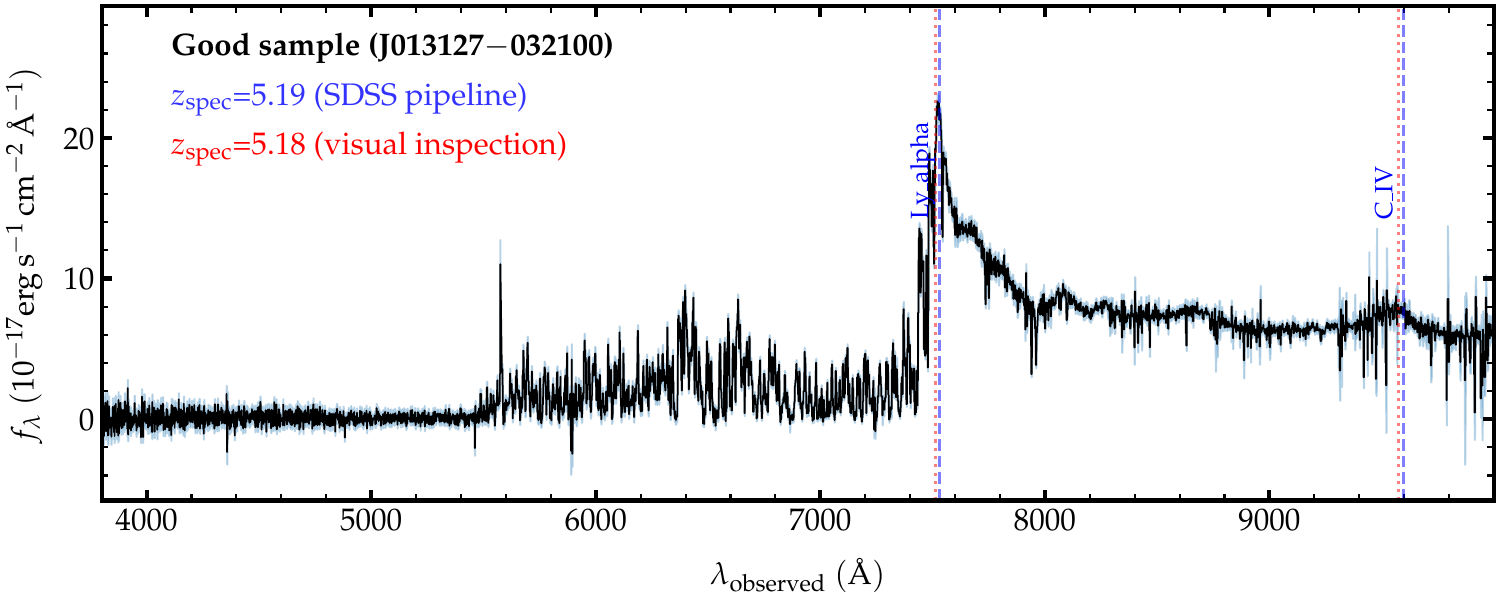}
\includegraphics[width=6.2cm]{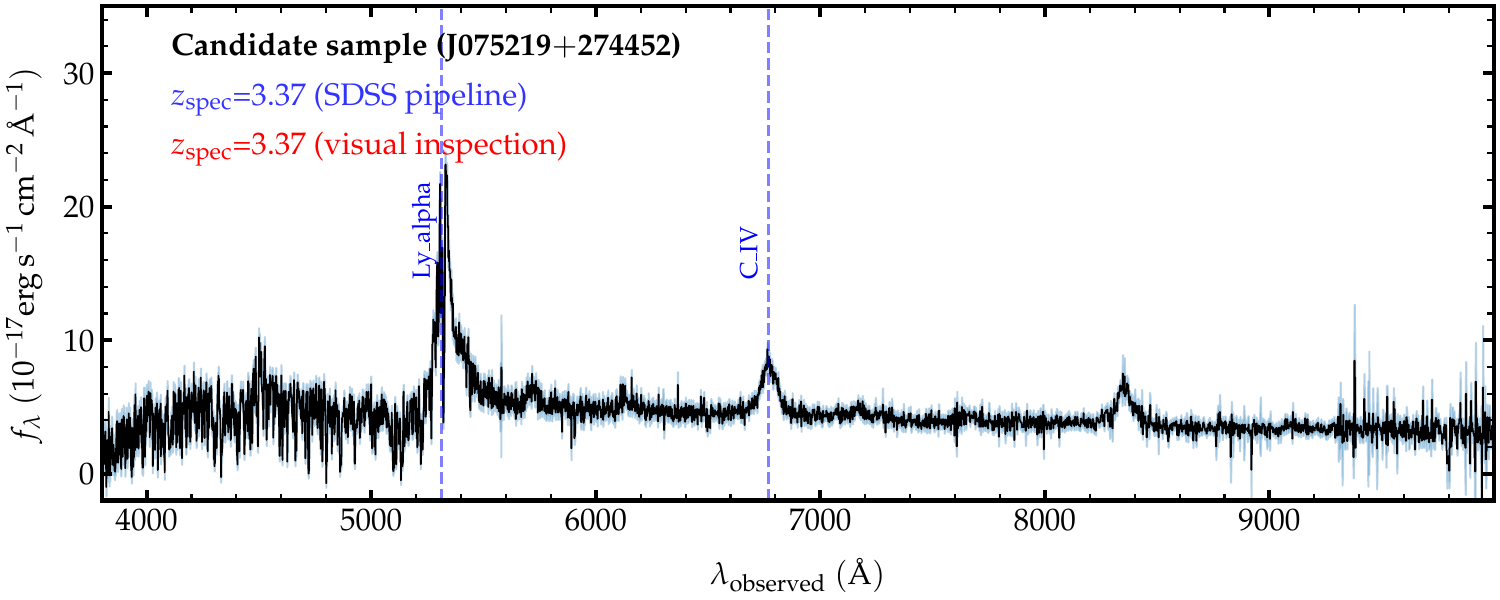}\\
\includegraphics[width=6.2cm]{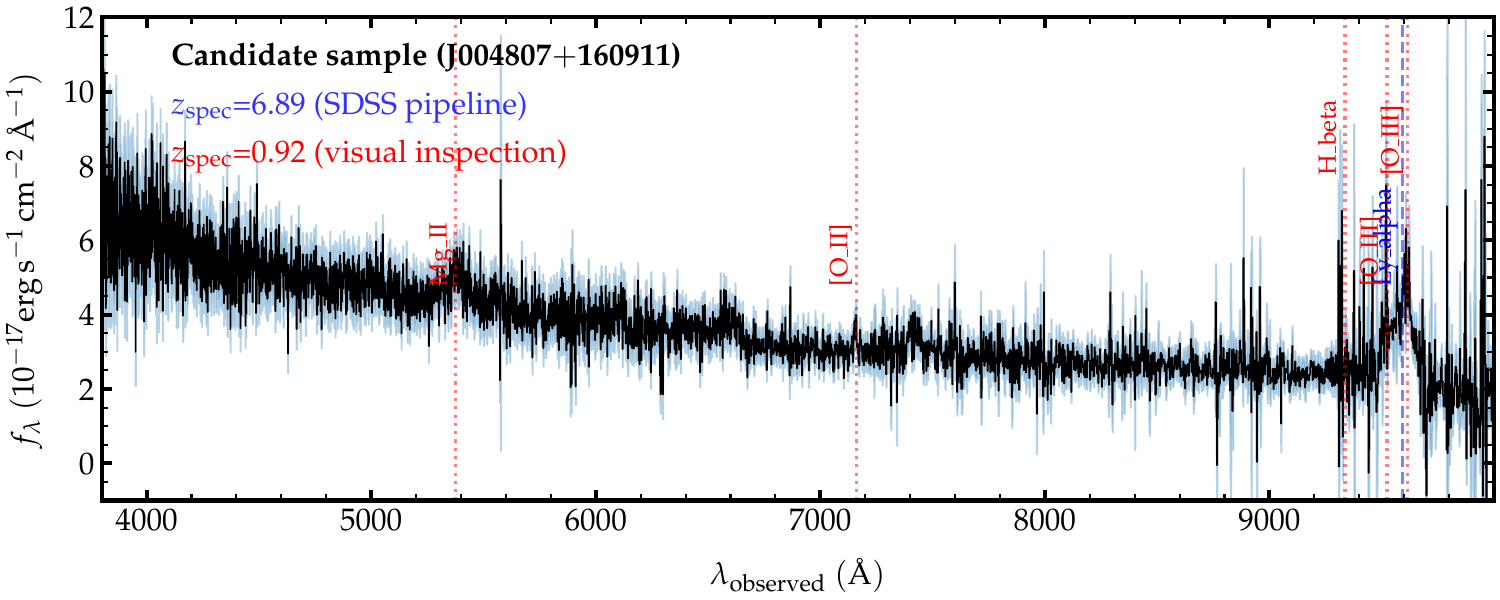}
\includegraphics[width=6.2cm]{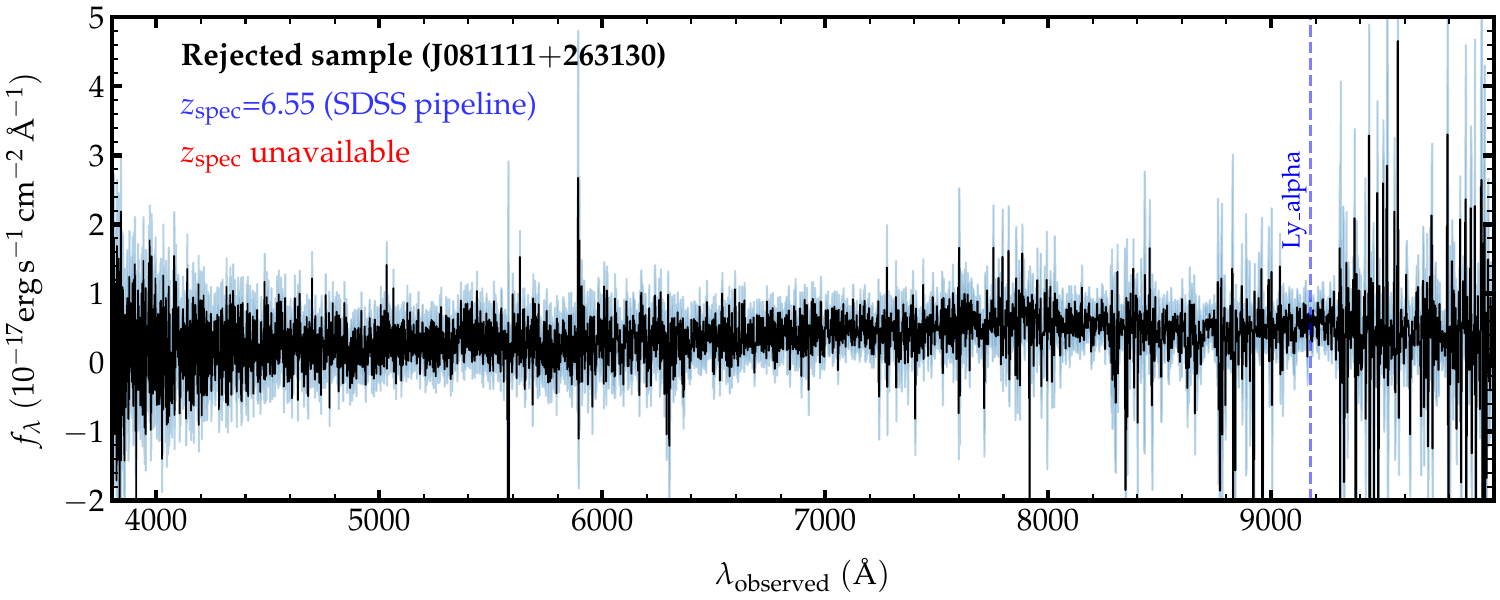}\\
\caption{Examples of SDSS spectra for different \texttt{z\_reliable\_flag} 
 values in our sample. The black lines show the observed SDSS spectra, with their $1\sigma$ uncertainties indicated in blue. Blue dashed and red dotted lines mark the expected wavelengths of prominent emission lines at the redshifts determined by the SDSS pipeline and our visual inspection, respectively. \textbf{Top left}: example source with good redshift (\texttt{z\_reliable\_flag = 1}) 
 showing agreement with our visual inspection; \textbf{top right}: example source with candidate high-redshift feature (\texttt{z\_reliable\_flag = 2}), 
 showing consistency with our visual inspection; \textbf{bottom left}: another example source with \texttt{z\_reliable\_flag = 2}, 
  but showing a discrepancy with our visual inspection; \textbf{bottom right}: example of unreliable redshifts with \texttt{z\_reliable\_flag = 3}, 
 where no predictable redshift can be made from our visual inspection.
\label{fig1:optical_spec}
}
\end{figure}

To select reliable redshifts for our high-redshift quasars, we performed a hierarchical flagging procedure similar to the tier-based method, introducing the \texttt{z\_reliable\_flag}. 
For \texttt{SOURCE\_Z} 
 with \texttt{VI} 
 or other reliable sources, we used \texttt{z\_reliable\_flag = 1}. 
 For a redshift derived from pipelines (\texttt{SOURCE\_Z = `PIPE'}), 
 we marked the robust pipeline redshifts (\texttt{ZWARNING = 0}) 
 with \texttt{z\_reliable\_flag = 2} 
 and wrong (unreliable) pipeline redshifts with \texttt{z\_reliable\_flag =  3}. 
The \texttt{z\_reliable\_flag} = 1 
 sources all had reliable redshifts according to visual inspection. 
The highest redshift in this category was \mbox{J1026+2542,} a well-known blazar with the first proper motion measurement of a VLBI radio jet \mbox{at {z > 5} \citep{2015MNRAS.446.2921F}.} 
Sources with \texttt{z\_reliable\_flag = 2} 
 might have had correct redshifts but required further verification before utilization. 
The \texttt{z\_reliable\_flag = 3} 
 sources were not recommended to be considered as RHRQs and were classified as ``rejected'' in our following analysis. Examples of SDSS spectra for different \texttt{z\_reliable\_flag} 
 values are demonstrated in Figure \ref{fig1:optical_spec}.

\begin{figure}[H]
\includegraphics[width=10.5 cm]{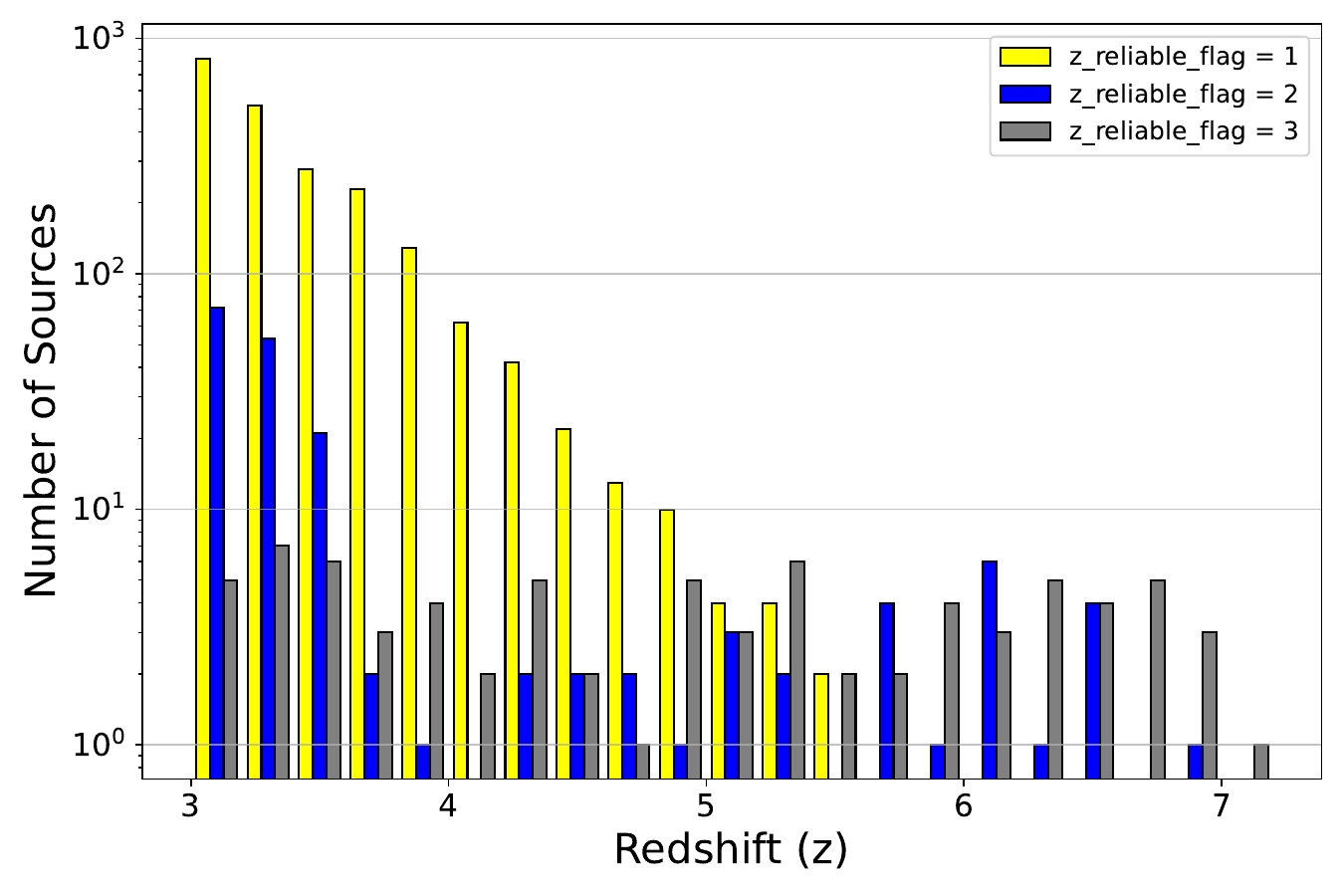}
\caption{The redshift distribution histogram of the matched catalog, with sources grouped by their redshift reliability flags.
\label{fig2:z_distribution}}
\end{figure}

The redshift distributions of sources with different reliability flags are presented in Figure~\ref{fig2:z_distribution}, where logarithmic binning is applied.
As seen in this figure, the number of sources with \texttt{z\_reliable\_flag = 1} 
 steadily decreases with an increasing redshift, in good agreement with the statistical trend observed in the original SDSS quasar catalog \citep{2020ApJS..250....8L}.
Sources with \texttt{z\_reliable\_flag = 2} 
 exhibit a similar trend at $z < 4$, but their counts slightly increase at higher redshifts. This suggests that misidentified redshifts become more prevalent at the high-redshift end, likely due to the limitations discussed earlier. 
Sources flagged as \texttt{z\_reliable\_flag = 3} 
 show an approximately flat distribution across all redshift bins, indicating a noise-like pattern with unreliable redshift estimates.
The distributions of sources from each redshift reliability flag are in good agreement with our expectation of source classifications.

\vspace{-10pt}

\subsubsection{Visual Inspection}
\begin{figure}[H]
\includegraphics[width=5.8 cm]{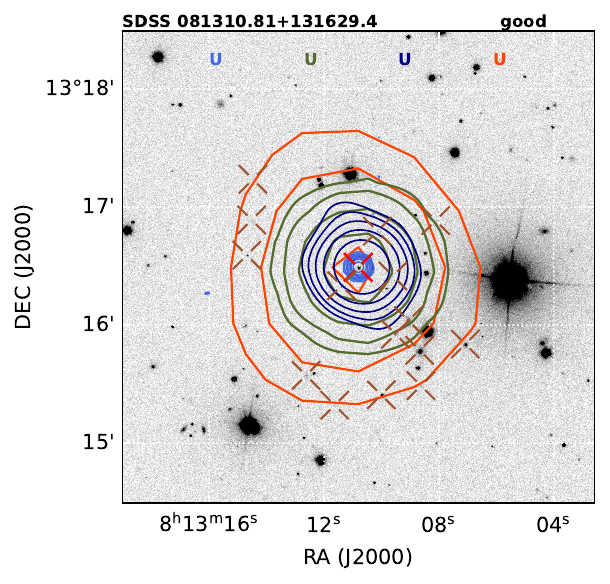}
\includegraphics[width=5.8 cm]{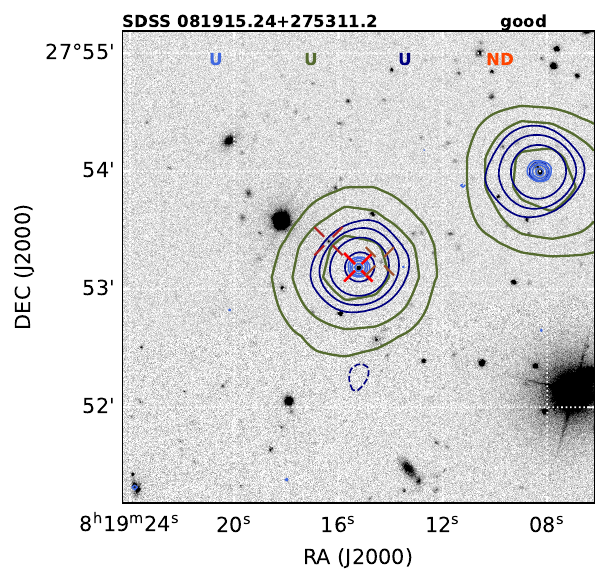}\\
\includegraphics[width=5.8 cm]{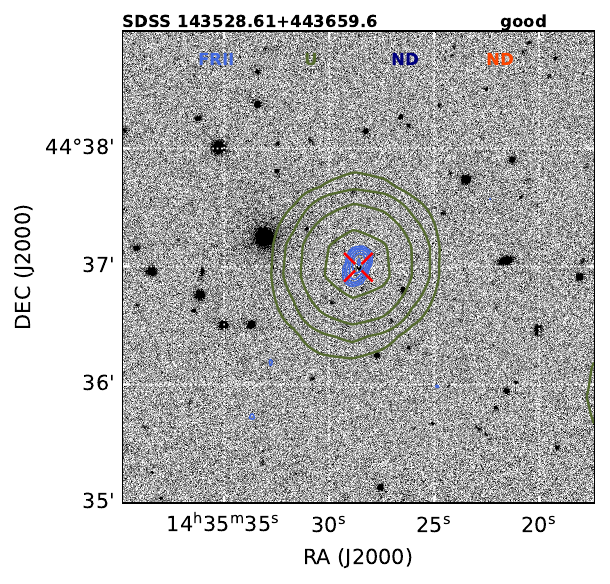}
\includegraphics[width=5.8 cm]{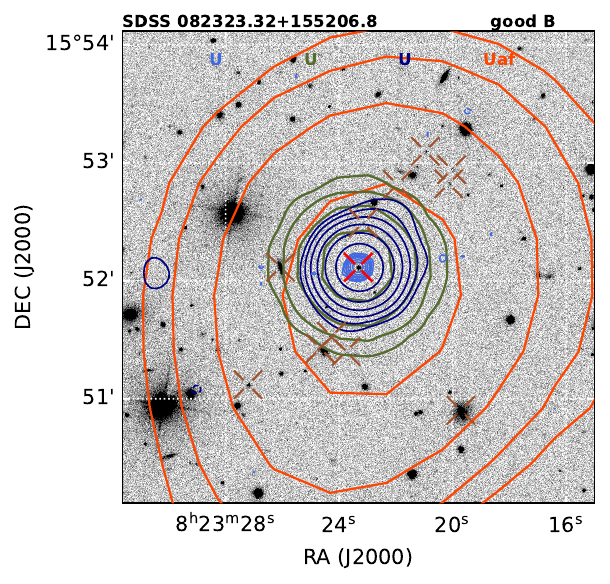}
\caption{Example
 overlay plots of the \textit{good}
 sample. In each panel, the background grayscale image is the optical SDSS image cutout, while the colored contours indicate the radio maps from different surveys (light blue: FIRST; green: NVSS; royal blue: RACS; orange-red: GLEAM).
The contours start at the $3\sigma$ level of each survey and increase by successive factors of 2. For weak sources with fewer than three contour levels, an additional contour at $\sqrt{2}$ times the base level is inserted between the first and second contours.
The inclined red cross at the image center marks the target HRQ from DR16Q. 
Brown crosses indicate low-redshift galaxies within the radio beam, and dark red crosses indicate low-redshift quasars within the beam.
At the top of each panel, the source identification codes from the respective radio catalogs are displayed, with font colors corresponding to the \mbox{contour colors.}
\label{fig:overlay_good}}
\end{figure}

\begin{figure}[H]
\includegraphics[width=5.9 cm]{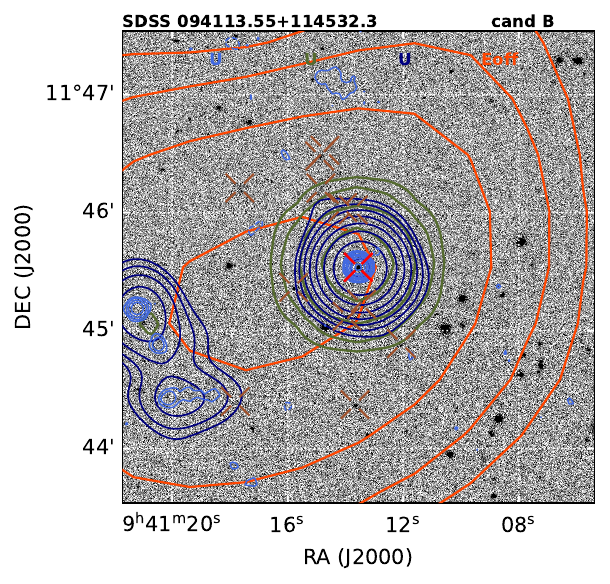}
\includegraphics[width=5.9 cm]{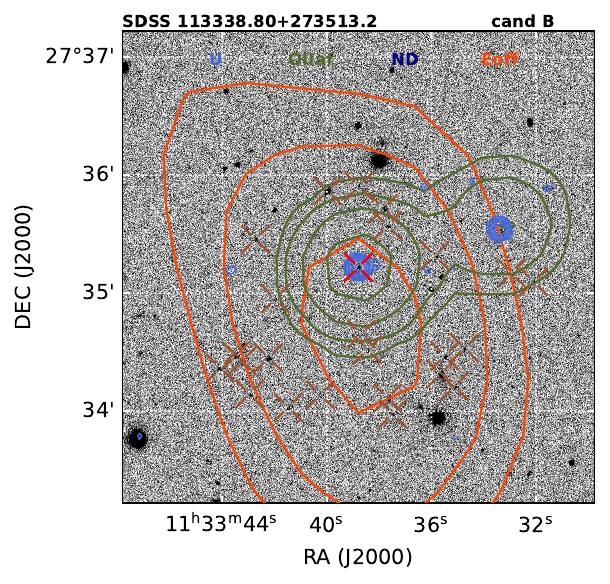} \\
\includegraphics[width=6.1 cm]{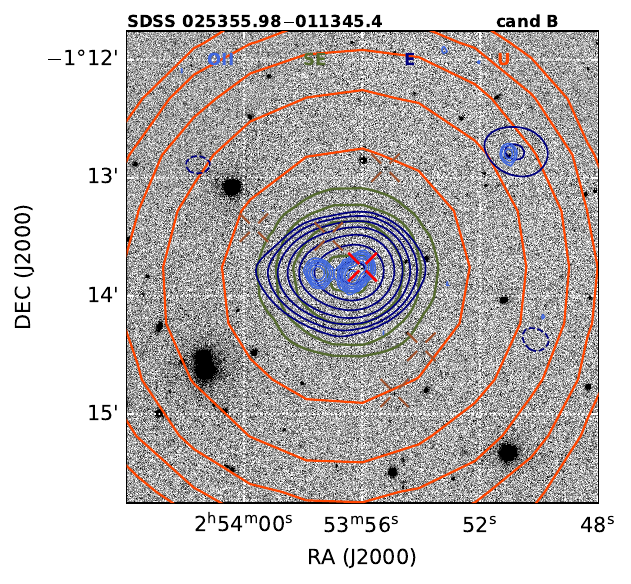}
\includegraphics[width=5.9 cm]{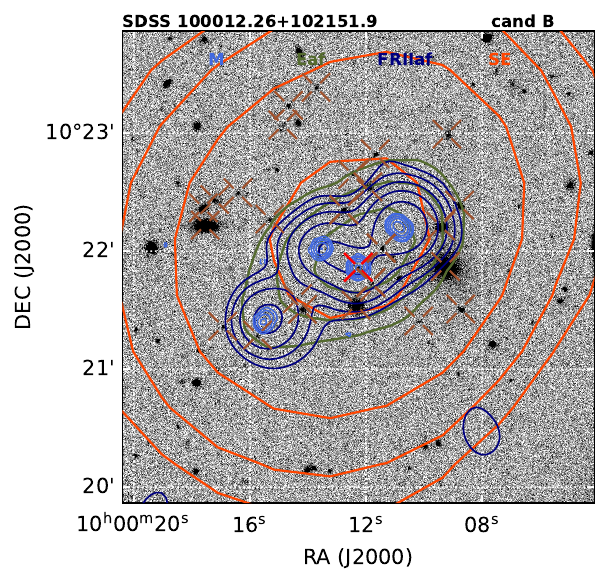} \\
\caption{Example 
 overlay images of the \textit{candidate} 
 sample.
The explanation of the plots is the same as for Figure~\ref{fig:overlay_good}.
\label{fig:overlay_cand}}
\end{figure}

\begin{figure}[H]

\includegraphics[width=5.8 cm]{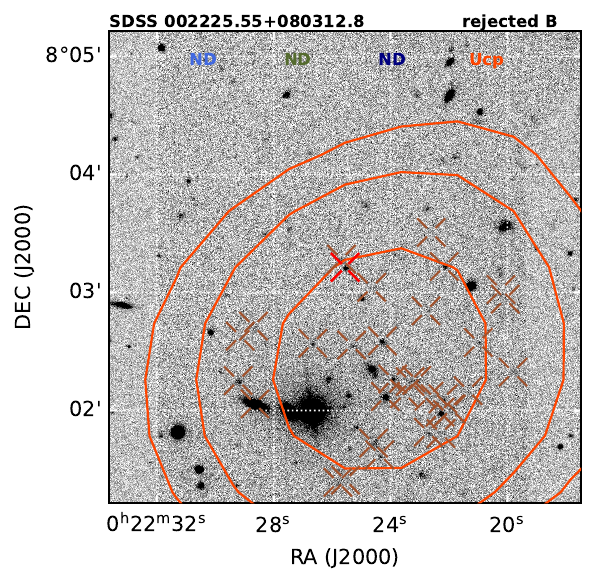}
\includegraphics[width=5.8 cm]{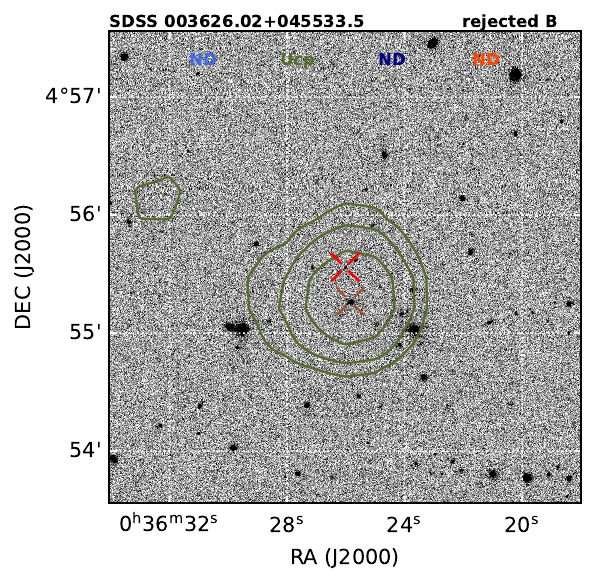} \\
\includegraphics[width=5.8 cm]{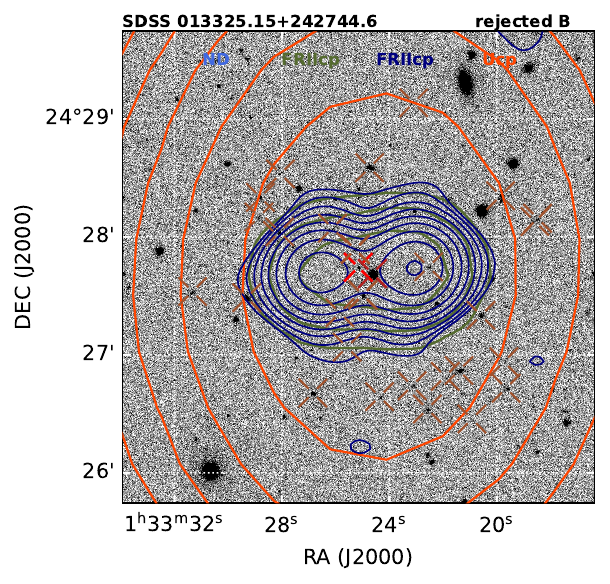}
\caption{Example 
 overlay images of the \textit{rejected} 
 sample.
The explanation of the plots is the same as for Figure~\ref{fig:overlay_good}.
\label{fig:overlay_bad}}

\end{figure}

For the matched high-$z$ quasar candidates (excluding the \texttt{z\_reliable\_flag = 3} 
 sources with unreliable redshifts), we made optical--radio overlay plots of each source based on the available data and performed a visual inspection for every source in our matched catalog.
In the overlay plots, the grayscale background is from the SDSS $i$-band image (or the $z$-band image if the $i$-band one is not useful). Radio images from each survey are overlaid with contours using different colors.
Images are centered at the target optical source, and all optical counterparts that are found within the radio beams are marked as inclined crosses (see examples in Figures~\ref{fig:overlay_good}--\ref{fig:overlay_bad}).

\vspace{-12pt}

During the visual inspection, we classified the radio structures of the RHRQ candidates into several morphological types according to their contour shapes. To facilitate description and identification, we applied a code-description approach similar to the identification work of \cite{2020ApJS..247...53K}. A set of codes was introduced to represent the source morphology and \mbox{types (see Table~\ref{tab5:ident_code}).}

\begin{table}[H] 
\caption{Radio identification codes and explanations.\label{tab5:ident_code}}
\scriptsize
\begin{tabularx}{\textwidth}{llL}
\toprule
\textbf{Code} & \textbf{Radio Morphology} & \textbf{Description} \\
\midrule
U & Unresolved & Single point-like component \\
SE & Slightly extended & Single component with slightly extended structure \\
E & Elongated & Single elongated profile that is likely distorted by nearby source flux densities \\
CJ & Core--jet & One circular component with a one-sided tail-like jet structure \\
FRI & Radio galaxy FR I & Radio galaxy with FR I-like morphology \\
FRII & Radio galaxy FRII & Radio galaxy with FR II-like morphology \\
OII & One-sided FR II & FR II-like radio source with optical core centered at one of the lobes \\
M & Multiple components & More than one component located within a small or continuous region \\
ND & Not detected & No radio emission detected at the position of the optical quasar \\
\midrule
\textbf{Code}
 & \textbf{Positional identification} 
 \textsuperscript{1} & \textbf{Description} 
 \\
\midrule
af & Affected & Radio emission affected by nearby sources or lobes \\
off & Off-centered & Radio source centroid shows a significant offset from the optical quasar \\
cp & Counterpart dominated & Radio source is centered on a low-redshift optical counterpart \\
\bottomrule
\end{tabularx}

\noindent{\footnotesize{\textsuperscript{1} If the optical quasar is not centered at the radio source, these additional codes are applied after the \mbox{morphology codes.}} }
\end{table}

Based on the categorization by visual inspection, each source was assigned to one of three main classes: the good sample (including good and good B), the candidate sample, and the rejected sample.

The reliable sample (\textit{good}) 
 refers to sources for which all associated radio surveys show contours well aligned with the high-$z$ optical quasar. These sources typically exhibit unresolved (U) radio structures consistent with the survey beams.
For a subset of the \textit{good} 
 sources, the radio contours from surveys with larger beams (e.g., GLEAM or NVSS) are slightly affected by nearby sources (`af'), 
 while the central position can still be reliably associated with the optical quasar. These are designated as \textit{good B}. 
 For such cases, the flux densities from NVSS or GLEAM may be slightly overestimated, but remain consistent within the quoted uncertainties.

Figure \ref{fig:overlay_good} demonstrates overlay plots of sources from the \textit{good} 
 sample, where a single unresolved component is detected in the available radio surveys and is well centered on the optical quasar despite the presence of background optical sources (top two panels).

The bottom left panel illustrates a classical Fanaroff--Riley type II (FR II) source \cite{1974MNRAS.167P..31F}, where the FIRST morphology shows a clear double-lobed structure, with the optical quasar located at the midpoint between the lobes. The bottom right panel presents an example of a \textit{good B} 
 source, where consistent associations are seen in FIRST, NVSS, and RACS, but the GLEAM detection is slightly affected due to its larger beam.

A candidate radio source is one whose radio emission is heavily affected or dominated by nearby radio lobes or low-redshift counterparts from at least one radio survey.
This kind of source often comes from a Tier~3 match from NVSS or GLEAM.
In most cases, the candidate source's FIRST counterpart (if one exists) is unresolved and in good alignment with the optical quasar. Its low-resolution radio structure is not centered at the target HRQ, but is rather dominated by or centered at nearby sources. We used the positional code `off' 
(Table~\ref{tab5:ident_code}) for radio sources with a large offset from the optical target or dominated by the blended radio emission of the target and nearby sources. The code `cp' 
is used to indicate if the radio emission is dominated by the nearby source.

For instance, in the example images from Figure~\ref{fig:overlay_cand}, the GLEAM detection in the top left panel is obviously affected by nearby radio sources. Its elongated GLEAM source is a combination of the target source as well as nearby radio sources, causing its flux density in GLEAM to be overestimated.
In the top right panel, aside from the off-centered GLEAM detection, the NVSS morphology appears FRII-like. While the FIRST detection confirms a compact unresolved core, its NVSS and GLEAM radio associations should be interpreted with caution.

In addition to the off-centered cases, we also found a certain number of the RHRQs to be showing FRI/II-like radio structures \cite{1974MNRAS.167P..31F}, while their optical counterparts were not always sitting between the radio lobes but rather at one of the lobe centers (marked with code OII, Table~\ref{tab5:ident_code}). Members of another small group show multiple unresolved FIRST components embedded within a single NVSS or GLEAM source (coded as M). 
These sources, if considered to be radio galaxies, could be relatively larger than typically expected for an HRQ.
We therefore classify these sources as \textit{candidates} since they could also be multiple unresolved radio sources or be caused by a chance alignment with a radio galaxy (RG) whose optical counterpart is not detected. 
The bottom two panels of Figure \ref{fig:overlay_cand} illustrate these cases: the bottom left panel shows an OII source, while the bottom right panel presents an M-type structure.

\begin{figure}[H]
\includegraphics[width=10.5 cm]{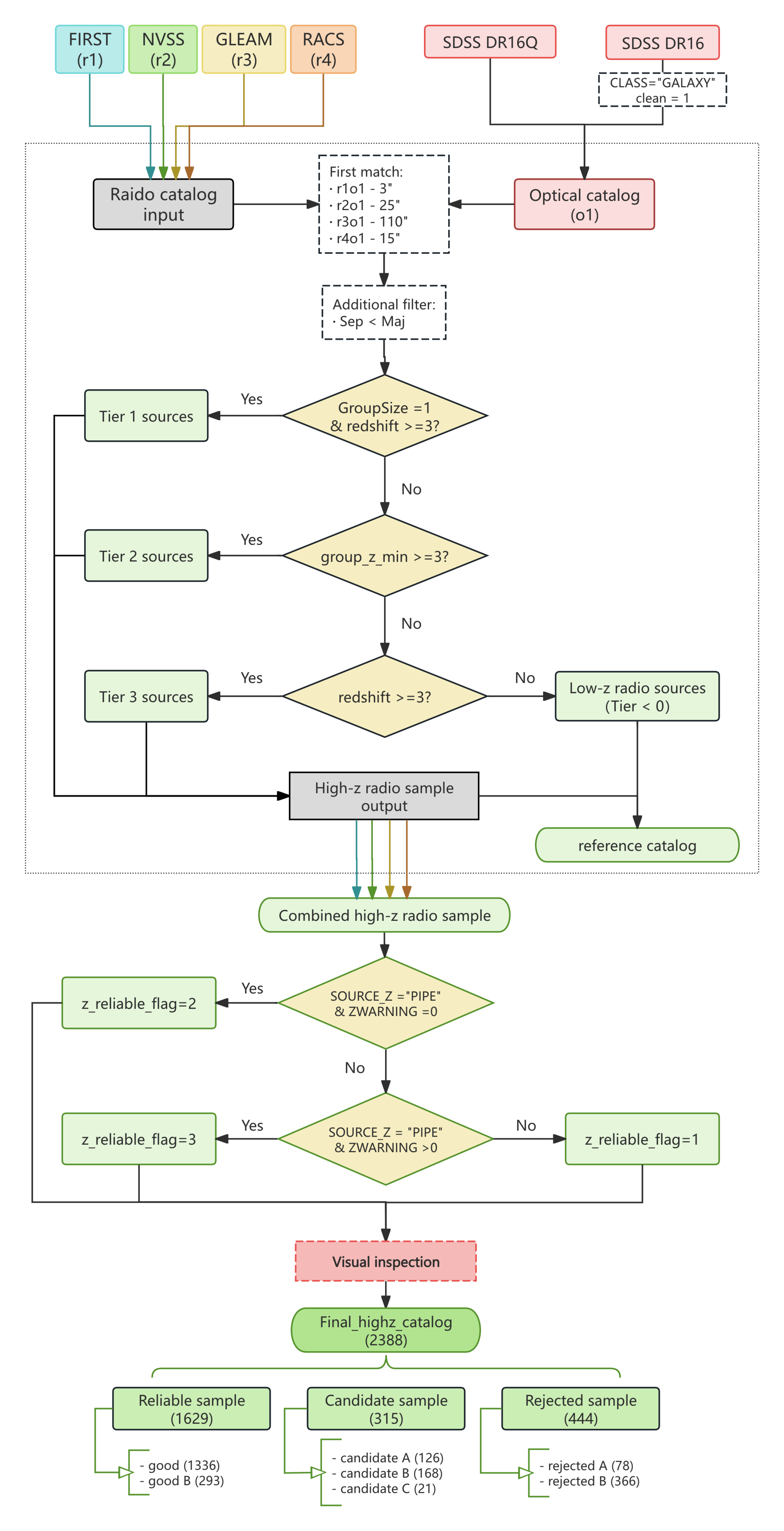}
\caption{The
 flowchart illustrating the optical--radio cross-matching procedure performed for constructing the high-redshift AGN sample.\label{fig1:flowchart}}
\end{figure}

A source will be \textit{rejected} 
if none of its radio detection shows good alignment with the optical quasar (e.g., mostly with code ND for high-resolution surveys and `off' or `cp' for low-resolution surveys). We note that this rejection from visual inspection only indicates that the radio flux density of the source is below the detection limits of the high-resolution radio surveys we use, which is not similar to the rejection based on unreliable redshifts.
Figure~\ref{fig:overlay_bad} displays three example plots of \textit{rejected 
} sources in our catalog.
The radio sources in the first two panels show the clear dominance of an optical source seen nearby (`cp'). For the last panel, a typical FRII galaxy coincident with a nearby optical galaxy is the preferred explanation for the origin of the radio emission, rather than the target HRQ.
The flowchart summarizing the cross-matching as well as the verification procedures is shown in Figure~\ref{fig1:flowchart}. 

\section{The Final RHzQCat}
\label{sec3:catalog}
The final high-$z$ catalog RHzQCat contains 2388 SDSS ``QSO'' sources obtained from the cross-matching procedure, with subsequent categorization after verification. By combining spectroscopic redshift validation and visual inspection, we defined three major subsamples:
\begin{itemize}
    \item Reliable sample. Sources with both trustworthy redshifts and a consistent radio morphology identified as good (including \textit{good} and \textit{good B}) during the inspection.
    \item Candidate sample. Divided into three subgroups:
    \begin{itemize}
        \item \textit{candidate A}: 
 sources with uncertain redshifts (\texttt{z\_reliable\_flag~$=2$}) 
 but visually classified as good;
        \item \textit{candidate B}: 
 sources with reliable redshifts but visually classified as candidate;
        \item \textit{candidate C}: 
 sources with inconsistencies in both verifications.
    \end{itemize}
    \item Rejected sample.
        \begin{itemize}
        \item \textit{rejected A}: 
 sources discarded based on unreliable redshifts without visual inspection;
        \item \textit{rejected B}: 
 sources rejected during the visual inspection.
        \end{itemize}
\end{itemize}

After the full cross-matching and inspection, the resulting RHzQCat consists of 1629 reliable sources (1336 \textit{good} 
 and 293 \textit{good B}), 
 315 candidates (126 \textit{candidate A}, 
 \mbox{168 \textit{candidate B},} 
 and 21 \textit{candidate C} 
 sources), and 444 rejected cases (78 \textit{rejected A} 
 and 366 \textit{rejected B}
 sources).
For clarity, we present two complementary tables derived from the same source list:
(1) a classification table, which records verification and morphological codes and parameters related to the verification (see Table~\ref{tabs1:class1}), and (2) a source parameter table, which lists physical properties such as redshift and flux densities (see Table~\ref{tabs1:parm2}).

\vspace{-4pt}
\begin{table}[H]
\caption{\label{tabs1:class1}The column names and descriptions for the classification table.}
\begin{adjustwidth}{-\extralength}{0cm}
\begin{tabularx}{\fulllength}{lllL}
\toprule
\textbf{Index} & \textbf{Col. Name} & \textbf{Unit} & \textbf{Description} \\
\midrule
1 & sdss\_name &  & Source name from SDSS DR16Q \\
2 & sdss\_ra & deg & Right ascension of the SDSS source \\
3 & sdss\_dec & deg & Declination of the SDSS source \\
4 & category &  & The final category of the source (\textit{good}, \textit{candidate}, 
 or \textit{rejected}) 
 \\
5 & sdss\_z &  & Redshift from DR16Q \\
6 & sdss\_source\_z &  & Source of redshift derived from DR16Q \\
7 & sdss\_zwarning &  & Binary flag on SDSS pipeline redshift estimate quality \\
8 & z\_reliable\_flag &  & Flag on the reliability of the target's redshift \\
9 & match\_tier\_code \textsuperscript{1} &  & The four-digit code showing the tier flags for each radio survey \\
10 & FIRST &  & Source name from the FIRST survey catalog \\
11 & NVSS &  & Source name from the NVSS survey catalog \\
12 & GLEAM &  & Source name from the GLEAM survey catalog \\
13 & RACS-DR1 &  & Source name from the RACS-DR1 survey catalog \\
14 & SEP\_FIRST & arcsec & The optical--radio separation from source positions in FIRST and SDSS \\
15 & SEP\_NVSS & arcsec & The optical--radio separation from source positions in NVSS and SDSS \\
16 & SEP\_RACS & arcsec & The optical--radio separation from source positions in RACS and SDSS \\
17 & SEP\_GLEAM & arcsec & The optical--radio separation from source positions in GLEAM and SDSS \\
18 & VI\_code\_FIRST &  & Radio identification code on source morphology and type on FIRST \\
19 & VI\_code\_NVSS &  & Radio identification code on source morphology and type on NVSS \\
20 & VI\_code\_RACS &  & Radio identification code on source morphology and type on RACS \\
21 & VI\_code\_GLEAM &  & Radio identification code on source morphology and type on GLEAM \\
22 & comment\_1 &  & Comment on the source identification from visual inspection \\
\bottomrule

\end{tabularx}
\end{adjustwidth}
	\noindent{\footnotesize{\textsuperscript{1} This is a combination of the matching tiers from different radio catalogs. Digits from left to right are tiers from FIRST, NVSS, RACS, and GLEAM, separately. For example, 1000 means a source only has FIRST detection with Tier 1; 0310 means the source has a Tier 3 detection from NVSS and a Tier 1 detection from RACS, but no detection in FIRST or GLEAM.}}

\end{table}

\begin{table}[H]
\caption{\label{tabs1:parm2}The column names and descriptions for the source parameter table.}
\begin{adjustwidth}{-\extralength}{0cm}
\begin{tabularx}{\fulllength}{lllL}
\toprule
\textbf{Index} & \textbf{Col. Name} & \textbf{Unit} & \textbf{Description} \\
\midrule
1 & sdss\_name &  & Source name from SDSS DR16Q \\
2 & sdss\_ra & deg & Right ascension of the SDSS source \\
3 & sdss\_dec & deg & Declination of the SDSS source \\
4 & category &  & The final category of the source (\textit{good}, \textit{candidate}, 
 or \textit{rejected}) 
 \\
5 & sdss\_z &  & Redshift from DR16Q \\
6 & FIRST &  & Source name from the FIRST survey catalog \\
7 & Fpeak\_first & mJy/beam & Source peak intensity at 1.4 GHz from FIRST \\
8 & Fint\_first & mJy & Integrated flux density of the FIRST source \\
9 & Rms\_first & mJy/beam & Rms noise of the FIRST image \\
10 & NVSS &  & Source name from the NVSS survey catalog \\
11 & S1.4\_nvss & mJy & Total flux density of the NVSS source at 1.4 GHz \\
12 & e\_S1.4\_nvss & mJy & Uncertainty of the flux density of the NVSS source \\
13 & RACS-DR1 &  & Source name from the RACS-DR1 survey catalog \\
14 & Fpk\_racs & mJy/beam & Peak intensity of the RACS-DR1 source \\
15 & e\_Fpk\_racs & mJy/beam & Error of the peak intensity of the RACS-DR1 source \\
16 & Ftot\_racs & mJy & Total flux density of the RACS-DR1 source \\
17 & e\_Ftot\_racs & mJy & Error of the total flux density of the RACS-DR1 source \\
18 & GLEAM &  & Source name from the GLEAM survey catalog \\
19 & Fpwide\_gleam & Jy/beam & Peak intensity of the GLEAM source in wide-band image \\
20 & e\_Fpwide\_gleam & Jy/beam & Fitting error on peak intensity in wide-band image of the GLEAM source \\
21 & Fintwide\_gleam & Jy & Integrated flux density in wide-band image of the GLEAM source \\
22 & e\_Fintwide\_gleam & Jy & Error of integrated flux density in wide-band image of the GLEAM source \\
\bottomrule

\end{tabularx}
\end{adjustwidth}
\end{table}

\section{Discussion} \label{sec4:discussion}
\subsection{Identified RHRQs Across Radio Surveys}
According to the numbers of identified RHRQs from different radio surveys, we can assess the ability to detect RHRQs using surveys with different resolutions.
In this part of the analysis, we did not include sources from the \textit{rejected A} 
 category where the redshifts are not reliable.

\begin{figure}[H]
\includegraphics[width=6.8 cm]{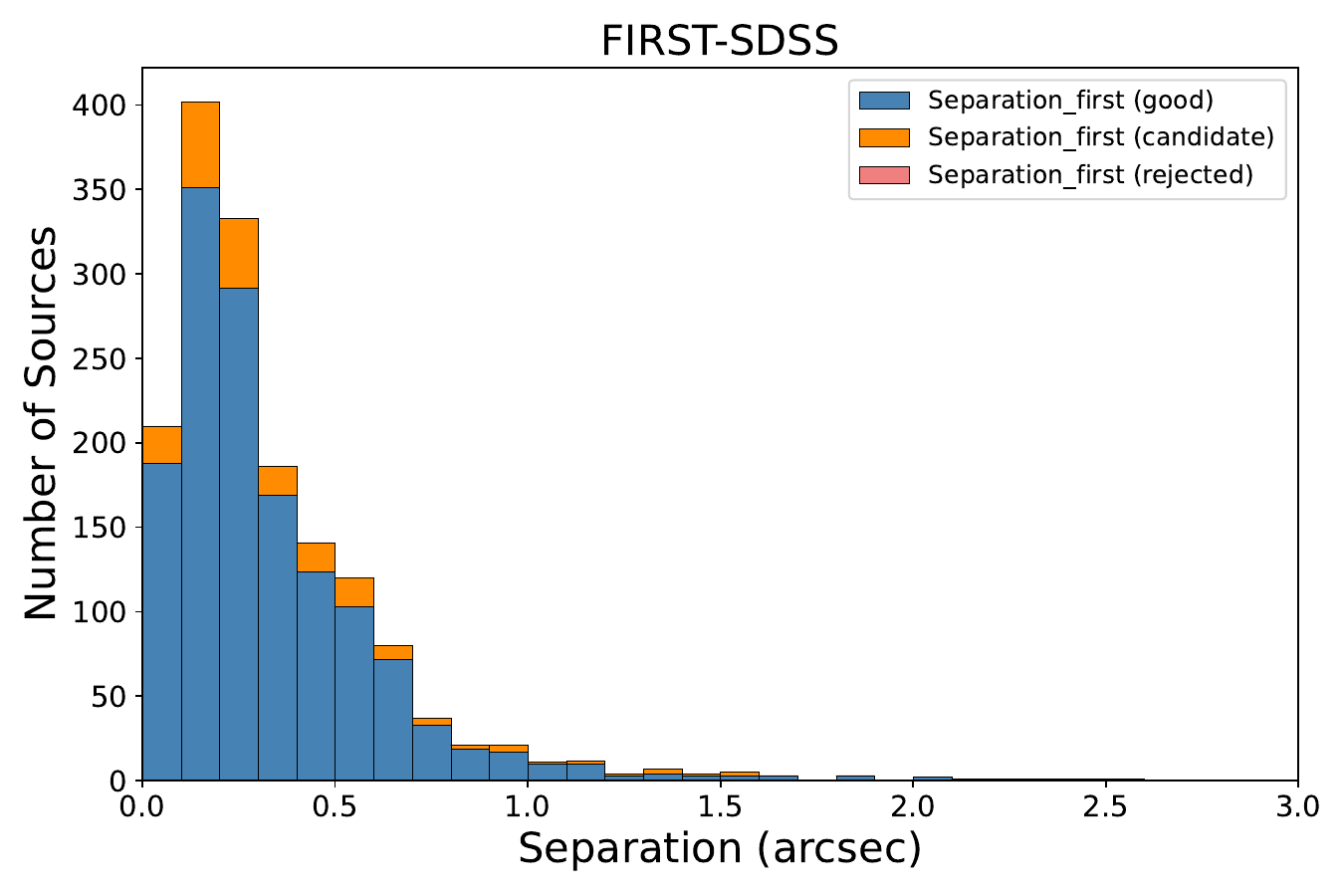}
\includegraphics[width=6.8 cm]{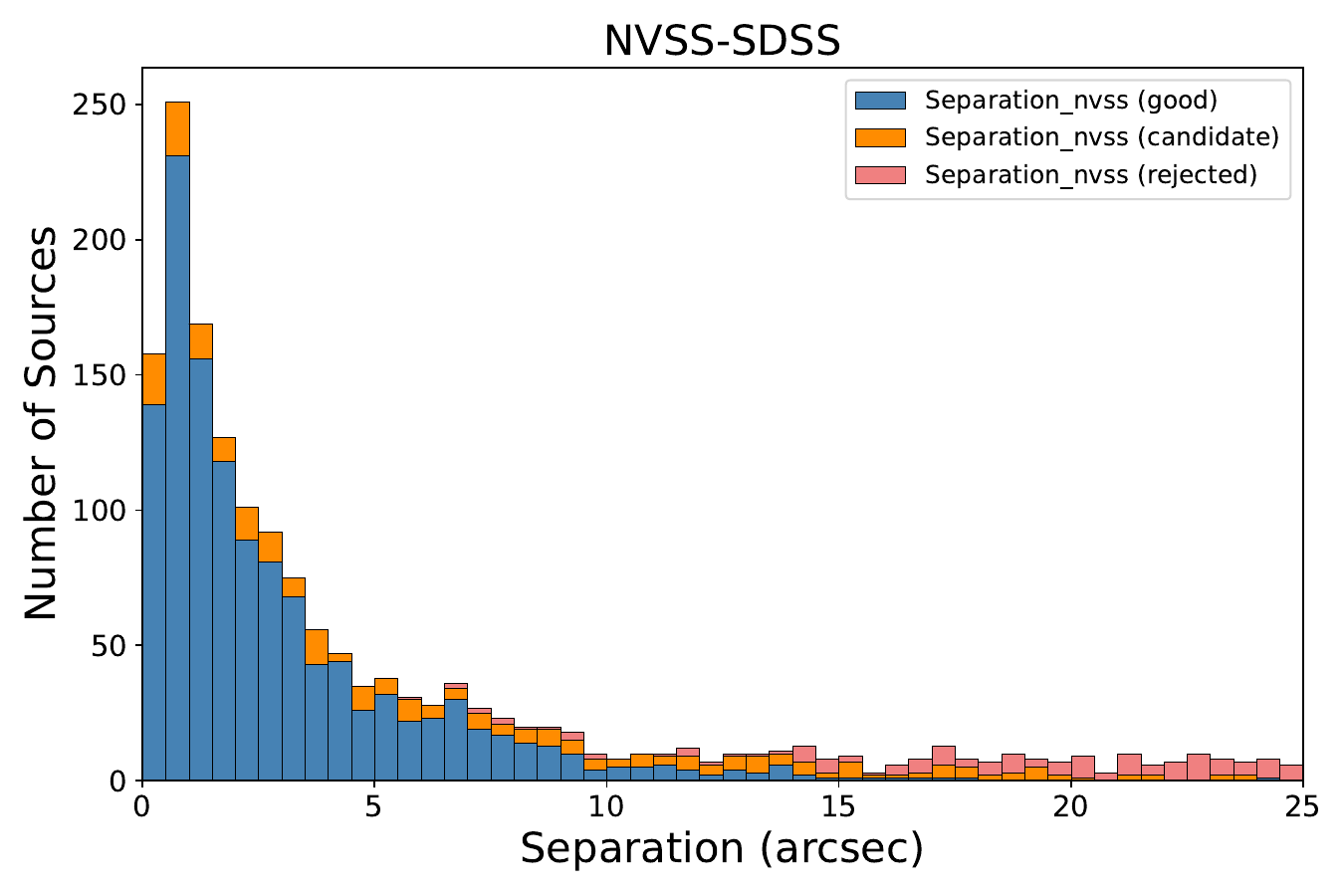} \\
\includegraphics[width=6.8 cm]{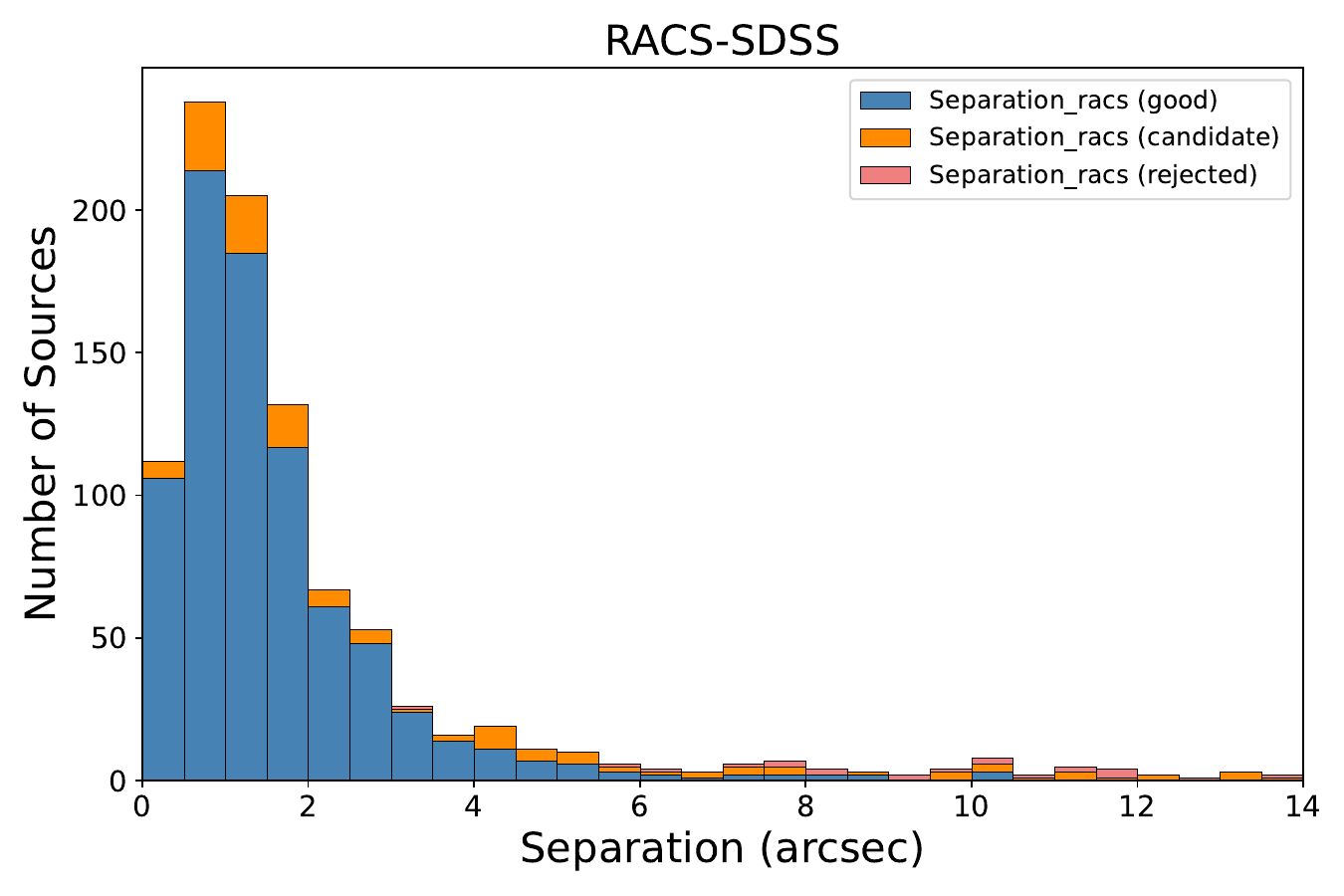}
\includegraphics[width=6.8 cm]{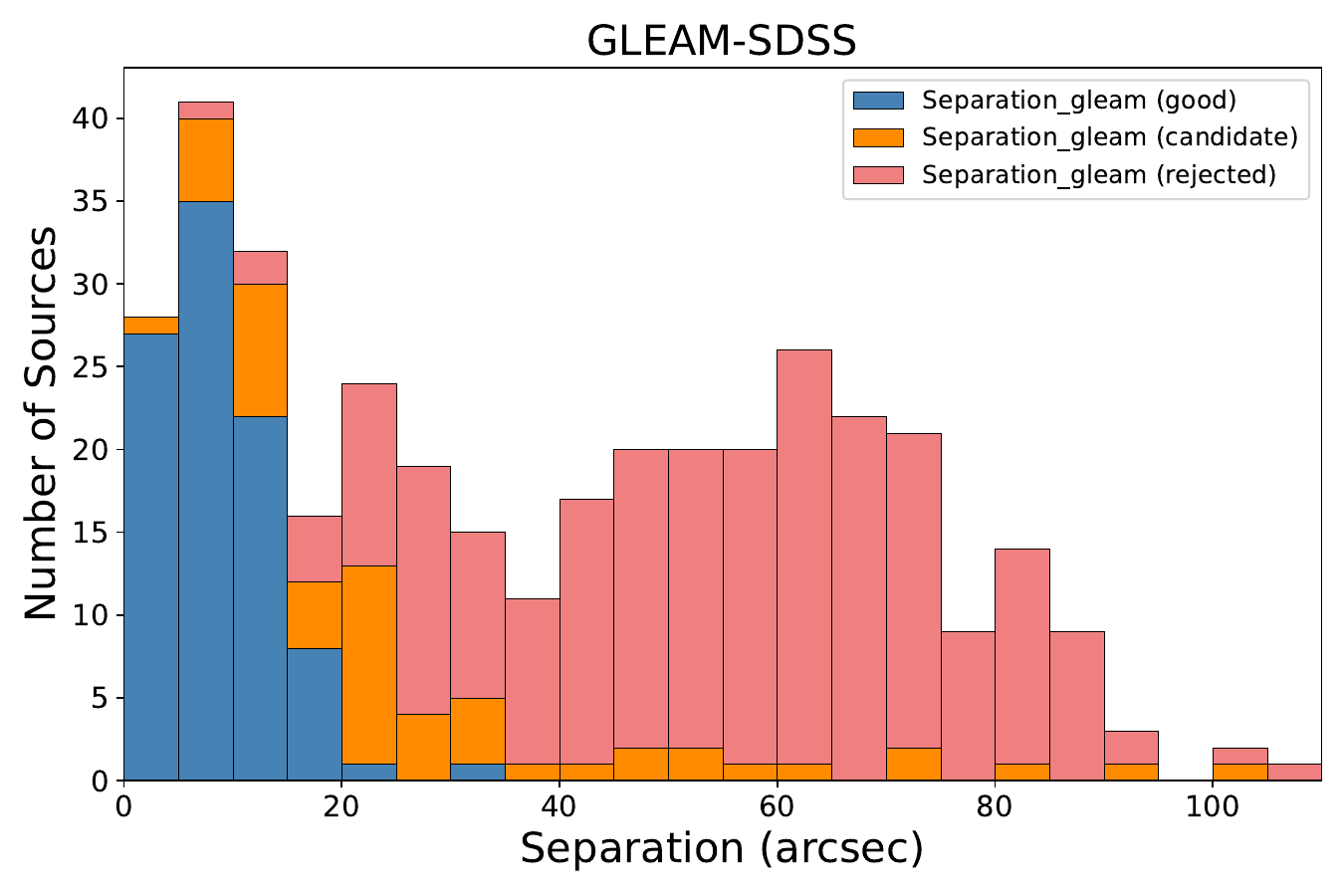}
\caption{The optical--radio separation distribution histograms in the compiled catalog for each radio survey used. \label{fig3:sep_distribution}}
\end{figure}

We list the number of sources of different categories detected from different surveys in Table~\ref{tab:detected_sources}.
Based on this table, we can conclude that the reliability increases with survey resolution. When going to GLEAM, where the beam size is large ($\sim$200$^{\prime\prime}$), many rejected sources emerge due to the contamination of the increased number of nearby sources. A detailed table with both tier combination and source categories are presented as Table~\ref{tabss:subgroup}.
We also present the optical--radio separation distribution for HRQs found in these radio surveys (Figure~\ref{fig3:sep_distribution}).
For FIRST and NVSS, the separation distribution in our RHRQ catalog is in good agreement with what is found in the literature for these surveys. The FIRST-SDSS separations show similar distributions where matching radii of >2.5$^{\prime\prime}$ are considered as a random association \cite{2002AJ....124.2364I,2002SPIE.4836..154K}. 
The NVSS-2dfGRS catalog \citep{1999PASA...16..247S} derived an optical--radio association separation of $\le$15$^{\prime\prime}$ through Monte Carlo tests where the majority of matched objects should be within $10^{\prime\prime}$.
For RACS and GLEAM, while no optical--radio associations were made here, 
our findings can be used as reference for such studies, e.g., searches involving RHRQs or even compact radio sources, in the future.

This analysis confirms that the optimal matching strategy and the resulting sample purity are intrinsically linked to the angular resolution of the radio survey. Consequently, the composite nature of our catalog, while providing unprecedented completeness, introduces a complex selection function that must be accounted for in subsequent statistical analyses.

\begin{table}[H] 
\caption{Number of categorized sources from each radio survey. \label{tab:detected_sources}}
\begin{tabularx}{\textwidth}{CCCCC}
\toprule
\textbf{Survey} & \textbf{Total} & \textbf{Reliable} & \textbf{Candidate} & \textbf{Rejected} \\
\midrule
FIRST & 1607 & 1406 (87\%) & 201 (13\%) & 0 (0\%) \\
NVSS & 1614 & 1218 (75\%) & 257 (16\%) & 139 (9\%) \\
RACS & 955 & 805 (84\%) & 129 (14\%) & 21 (2\%) \\
GLEAM & 370 & 91 (24\%) & 54 (15\%) & 225 (61\%) \\
\bottomrule
\end{tabularx}

\end{table}

\vspace{-13pt}

\subsection{Radio Properties in the Compiled Catalog}

\subsubsection{Radio Morphology}

In the RHzQCat, the majority of the sources show a single unresolved structure.
Among the reliable sources, compact objects
with morphological types of U or Uaf constitute $\sim$95\% of the total sample. 
Compared with a compact source rate of $\sim$90\% at low redshifts \citep{2020ApJS..247...53K}, our results demonstrate the compact nature of HRQs.
This pronounced dominance of compact morphology at high redshifts could be attributed to a combination of selection, physical, and cosmological effects, although the relative importance of each remains unclear.
For example, according to the orientation-based unified model \cite{1995PASP..107..803U}, AGNs with smaller jet inclination angles toward the line of sight (e.g., blazars) appear visually compact and are more easily detected due to Doppler boosting compared with larger-inclination objects at the same redshift, whose radio structures are more extended.
At early cosmological epochs, radio jets are also expected to be younger and to evolve within denser galactic environments, naturally favoring a compact steep-spectrum (CSS), a gigahertz-peaked spectrum (GPS) \citep{2021A&ARv..29....3O}, or compact symmetric objects (CSOs) \citep{1994ApJ...432L..87W,2024ApJ...961..240K}.
In addition, the surface brightness of extended radio lobes is thought to be suppressed by inverse-Compton scattering on Cosmic Microwave Background (CMB) photons at high redshifts, with a scaling factor \mbox{of $(1+z)^{-4}$ \citep{2017MNRAS.468..109W}.}

For extended sources, given the predominantly compact nature of HRQs and the resolution limits of the radio surveys used, we identified a source as an RG or RG candidate only from its extended structure (E, CJ, or FRI/II) from FIRST or RACS. 
This yielded \mbox{64 RGs} within the \textit{good} 
 sample and 46 RG candidates within the \textit{candidate} 
 sample. The RG candidates typically exhibit FR-like structures but their optical counterparts are located at one of their lobes, showing a core--jet (CJ) or one-sided FR II (OII) morphology. There are also a few M-shaped radio sources that exhibit multiple unresolved components (from FIRST) within a complex radio structure from NVSS or RACS.
The one-sided and M-shaped radio galaxy candidates could be a mix of single unresolved radio sources and chance alignments of the HRQ with a foreground radio galaxy without an optical counterpart. In the former case, only the FIRST detection is reliable; in the latter, the corresponding HRQ is not a \mbox{radio emitter.}

Radio galaxies at a high redshift are of particular astrophysical importance. Their largest linear sizes provide direct constraints on the evolutionary stages of AGN jets in the early Universe, while correlations with spectral properties offer insights into particle acceleration and energy losses.
Moreover, the size distribution and morphology of high-$z$ RGs encode valuable information on the physical conditions of the intergalactic medium (IGM), such as density, pressure, and magnetic field strength.
Further investigation of these RG candidates---through multi-wavelength observations and higher-resolution radio imaging---will therefore be crucial for confirming their nature and exploiting their potential as probes of AGN evolution and the high-redshift IGM.

We also searched for possible dual AGNs, and only one reported AGN pair was found at $z \sim 3$, namely J1307+0422 \cite{2018MNRAS.477..780V}.
We started this by looking for matched groups with more than one $z \ge 3$ object among the Tier~2 and 3 sources.
The candidate groups are listed in Table~\ref{tabs3:dual_candidates}, where all of the objects are \textit{rejected}, 
 except for the known pair J1307+0422.
This can be expected since most dual AGNs tend to be weak at radio frequencies \citep{2015ApJ...799...72F,2024ApJ...969...36X}, which makes it even harder to find such sources from shallow surveys at redshifts above 3.
On the other hand, the finding of J1307+0422 indicates that our method is applicable for the search for dual AGN systems. Future studies using deeper surveys could be promising to reveal more AGN pairs and dual AGNs.

\subsubsection{Flux Density and Luminosity}
Since most of the sources in RHzQCat have flux densities measured at 1.4 GHz, we calculated their monochromatic radio luminosity. We took sources from the \textit{reliable} 
 sample as well as the \textit{candidate B} 
 sample (these sources have valid NVSS or FIRST detections).

\begin{figure}[H]
\includegraphics[width=\textwidth]{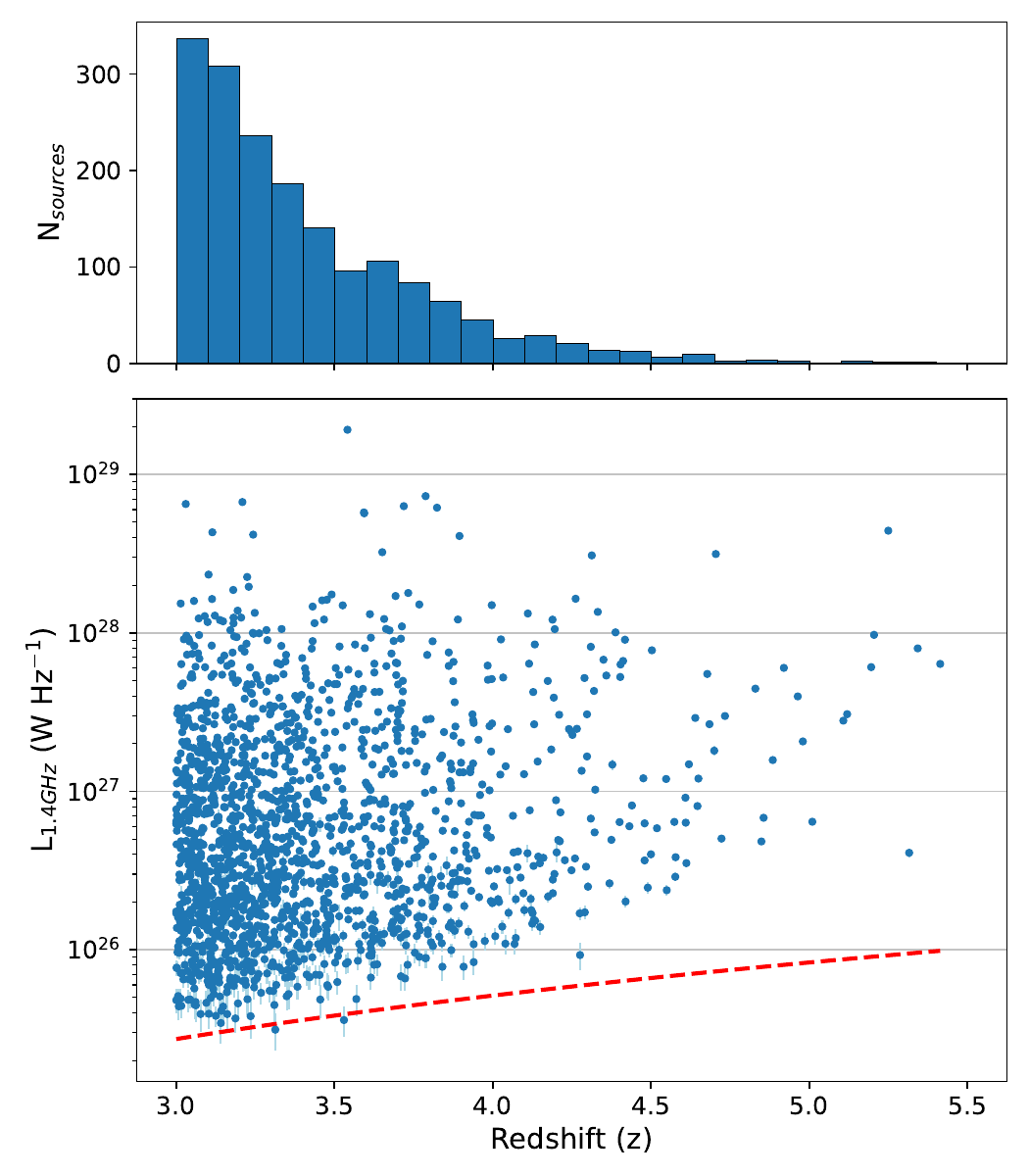}
\caption{Source
 numbers and monochromatic rest-frame 1.4 GHz radio luminosities of the sources in the compiled catalog. \textbf{Upper panel}: source distribution histogram vs. redshift; \textbf{lower panel}: radio luminosity at 1.4 GHz as a function of redshift. Blue dots represent \textit{good} sources with FIRST or NVSS detections. The red dashed line indicates a $3\sigma$ detection threshold of $\approx$0.5~mJy from FIRST. \label{fig:f_vs_l}}
\end{figure}

The rest-frame 1.4 GHz radio luminosity (L$_{\mathrm{1.4GHz}}$) was calculated from the observed flux density (taken from FIRST, or from NVSS if there was no FIRST detection) using the following standard formula:
\begin{equation}
     \mathrm{L}_{\nu,\mathrm{rest}} = 4\pi \mathrm{D_L}^2 \mathrm{S}_{\nu} (1+z)^{\alpha-1} \,\, [\mathrm{W}\,\mathrm{Hz}^{-1}],
\end{equation}
where D$_{\rm L}$ is the luminosity distance calculated from the cosmological parameters, and $\alpha$ is the radio spectral index defined as S$_{\nu} \propto \nu^{-\alpha}$.
We adopted a characteristic value of \mbox{$\alpha$ = 0.75} for the calculation \citep{2018ApJS..239...33Y}.
We note that for core-dominated flat-spectrum sources ($\alpha \approx 0$ at high redshifts), this may lead to an overestimation of the K-correction by about $20\%$, e.g., $(1+z)^{-1}$ for $z$ = 4.
A histogram showing the source distribution with redshift and a plot of the radio luminosities are displayed in Figure~\ref{fig:f_vs_l}.

Sources with reliable 1.4 GHz measurements in the RHzQCat populate the redshift range from 3 to 5.3, and have a wide range of radio luminosities L$_{\mathrm{1.4GHz}}$ from \mbox{$10^{25.5}$ to $10^{29.3}$~W\,Hz$^{-1}$.}
The luminosity distribution is shown in the bottom panel of \mbox{Figure~\ref{fig:f_vs_l},} 
clearly illustrating a well-known degeneracy in flux density-limited surveys: we can detect high-luminosity sources at very high redshifts and lower-luminosity sources at closer distances. This creates the diagonal lower envelope in the luminosity--redshift plot, which is primarily defined by the survey flux density limit.

The data are in good agreement with the $3\sigma$ detection threshold of FIRST ($\approx$0.5~mJy), and the luminosity values span nearly four orders of magnitude. At the highest redshifts, the number of RHRQs becomes smaller and the luminosity range is confined \mbox{to~$(10^{26}$ -- $10^{28})$~W\,Hz$^{-1}$.}

\begin{figure}[H]
\includegraphics[width=\textwidth]{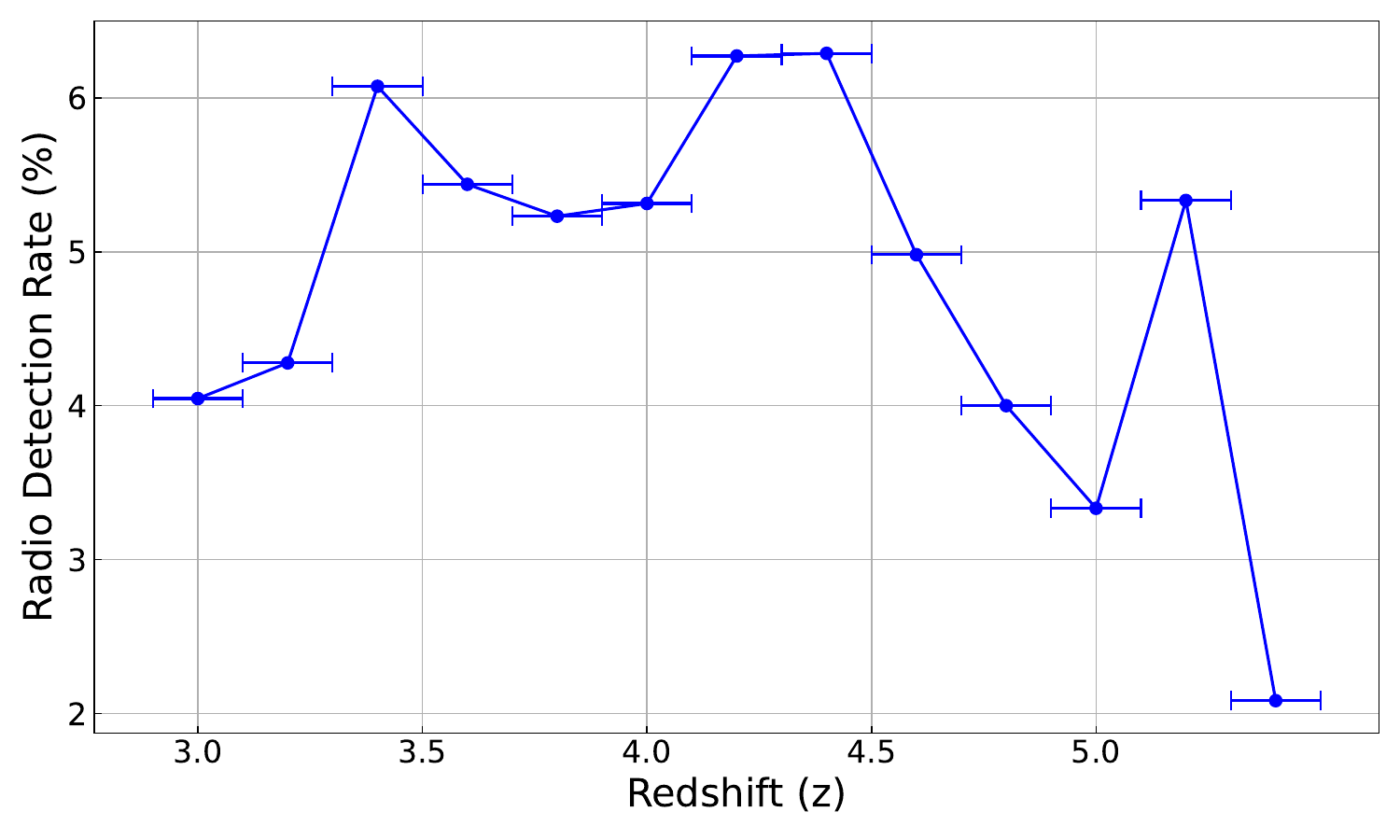}
\caption{The
 radio detection rate of HRQs estimated from RHzQCat. \label{fig:highz_rate}}
\end{figure}

We also plot the detection rate distribution ($\mathrm{N}_{\mathrm{radio}} / \mathrm{N}_{\mathrm{optical}}$, where the number of optical quasars $\mathrm{N}_\mathrm{optical}$ are from DR16Q and the number of radio quasars $\mathrm{N}_\mathrm{radio}$ from the \textit{good} and \textit{candidate A} source in our catalog) as a function of redshift in Figure~\ref{fig:highz_rate}.
This may indicate a higher activity period of RLAGN at $z \sim 4$. This result is in alignment with previous studies \citep{2015MNRAS.446.2483S} suggesting that at around redshift 4, the growth of SMBHs becomes faster.

\vspace{-6pt}

\subsection{Comparison with Previous RHRQ Sample}

P17 \cite{2017FrASS...4....9P} focused on RHRQs at $z\ge4$ but was not limited to SDSS, FIRST, and NVSS.
To make a fair comparison, we defined a subgroup of the P17 sample restricted to the SDSS footprint and radio detections (FIRST or NVSS), and cross-matched it with our catalog.
Of the 116 sources from the P17 subgroup, we found no SDSS source with available FIRST or NVSS detection that was missing from our catalog. (Nine sources had SDSS names in P17 and were not present in our catalog, but these were not found in DR16Q either.)
In addition, among the overlapping objects, two \textit{reliable} 
 sources (SDSS~001357.16 $-$ 083334.7 and SDSS~092554.13+194349.8) that showed FIRST upper limits in P17 had available NVSS and GLEAM flux densities in our catalog, indicating these two sources are extended RG candidates.
We also found 43 sources (13 \textit{good} 
 and 30 \textit{candidates}) 
 not in P17, mostly thanks to the larger matching radius we applied.

For sources at redshifts 3 -- 4, the largest radio quasar sample known to date is \mbox{from \cite{2021MNRAS.508.2798S},} where 102 $z\ge3$ quasars were selected with flux density S$_{\mathrm{1.4 GHz}} \ge 100$~mJy, whose redshifts were collected from the NASA/IPAC Extragalactic Database (NED)\endnote{\url{https://ned.ipac.caltech.edu} 
}.
Our catalog made full use of the SDSS redshifts and boosted the number of $z\ge3$ radio quasars to 1636 (plus 308 candidates), significantly increasing the number of RHRQs.

In recent years, \cite{2024A&A...689A.174C} selected 29 high-$z$ radio quasars and performed radio spectral analysis of these sources. In our catalog, three of their sources were rejected due to unreliable redshifts (0216+05, 1153+07, and 1309+03) from SDSS, and four sources (0011+14, 0115+03, 0949+23, and 1458+17) were categorized as \textit{candidate B} 
 because their GLEAM flux densities do not correspond to the expected high-$z$ sources. This means that their GLEAM flux densities are obviously overestimated and their original low-frequency spectra would be flatter or of a MHz-peaked shape.

\subsection{Sample Biases and Limitations}
The potential biases in our compiled catalog can be considered from two aspects: \mbox{(1) whether} we are missing HRQs that should have been detected in the radio surveys and (2) whether we are including spurious associations that are not genuine HRQs. Concerning the first aspect, our cross-matching validation with P17 indicates that no radio sources are missing at $z \geq 4$, at least within the SDSS--FIRST and SDSS--NVSS footprints. Given the relatively large matching radii and the tier-based flagging applied during cross-matching, we believe that the number of missing sources was minimized in our RHzQCat.

For the second aspect, the possibility of chance alignments remains.
Specifically, a target HRQ may be incorrectly associated with a radio lobe from a foreground low-redshift galaxy or from an optically faint quasar.
This can arise for two reasons:
\begin{enumerate}
    \item The optical parent catalog, although based on SDSS DR16, does not fully include galaxies that may be classified as STAR or those that fall outside the \texttt{clean = 1} 
 selection. To avoid an excessive catalog size and unnecessary visual inspection, we did not incorporate these sources in the main cross-matching.
    \item Although our optical--radio matching allows for slightly larger positional offsets than in some previous studies, chance alignments can occur at any offset if a radio lobe happens to coincide with the optical position.
\end{enumerate}

Cases of chance alignments are particularly likely for orphan detection in NVSS or GLEAM samples with an extended or distorted morphology that deviates from the compact unresolved structures expected for true radio quasars (e.g., sources in the \textit{good B} 
 or \textit{candidate} 
 categories).

It is important to acknowledge a fundamental limitation of our catalog: it is not a statistically complete, flux density-limited sample. The heterogeneous depths and frequencies of the radio surveys used (FIRST, NVSS, RACS, and GLEAM) imply that the RHzQCat is not characterized by a single flux density threshold and is not equally sensitive to different spectral types.
As a result, the catalog is not directly suitable for deriving luminosity functions or source counts without careful modeling of the underlying selection function, which lies beyond the scope of this paper.

\subsection{Future Prospects}
The RHzQCat provides immense value as the largest and most robust compilation of radio-detected quasars at $z \sim$ 3 -- 5 to date. Its primary strength lies in its role as a foundational resource for subsequent targeted studies that require a large, vetted sample of high-redshift radio-loud AGNs. Rather than for statistical population studies, this catalog is ideally suited for the following:
\begin{itemize}
    \item High-resolution follow-up observations. Bright radio sources from our reliable sample act as a prioritized target list for high-resolution VLBI imaging studies. The privilege of VLBI imaging and parameterizing of HRQs allow studies of the innermost AGN cores as well as pc-scale jets. This will help us in (1) measuring proper motions and estimating  apparent jet speeds to understand jet kinematics and energy transport in the early Universe, (2) constraining the brightness temperature to distinguish between Doppler-beamed and truly young, compact sources, and (3) precisely registering the radio core against the optical position to within milliarcsecond accuracy. Recent investigations of radio quasars at $z \sim$ 3 -- 4 \cite{2024Univ...10...97F,2025Univ...11...91G,2025Univ...11..157K} have demonstrated the crucial role of such a catalog in follow-up studies.

    \item Radio galaxy studies at high redshifts. High-redshift radio galaxies (HzRGs) are among the rarest and most intriguing AGNs in the early Universe. These sources are believed to be the most massive galaxies, and are rich in dust and gas \citep{1998MNRAS.301L..15B,2018MNRAS.480.2733S}. Studies of HzRGs and their environment at various redshift ranges provide insights into the assembly and evolution of these large and massive objects at different stages. To date, the most distant HzRGs known are at $z\approx5.5$ \cite{2020PASA...37...26D}, while the number of HzRGs at $z>3$ is still very limited \citep{2019MNRAS.489.5053S,2020AJ....160...60Y,2025A&A...697A.238C}. In the RHzQCat, we identified 110 RGs or RG candidates, about $5\%$ of the total number of sources.
    
    \item SKA pathfinder studies. This work serves as a critical testbed for the SKA era. Our method makes use of the catalogs (RACS-DR1 and GLEAM) from SKA precursor and pathfinder telescopes (i.e., the Australian Square Kilometer Array Pathfinder, ASKAP, and the Murchison Widefield Array, MWA). The RHzQCat and the matching procedure we developed can serve as a benchmark for developing and testing source-finding and cross-matching techniques for future deep, large-area radio surveys from SKA1-LOW.
    The multi-survey, multi-resolution cross-matching methodology developed here is directly applicable to the vast and complex datasets that SKA will produce. The faint end of our sample hints at the immense population of radio-weak and low-luminosity AGNs that SKA will uncover, allowing us to trace AGN activity and feedback down to previously inaccessible levels at these redshifts. In parallel, the upcoming Large Synoptic Survey Telescope (LSST) \cite{2014IAUS..304...11I,2019ApJ...873..111I} will provide an unprecedented census of HRQs with deep optical photometry and variability-based AGN identification, which will enable a far more complete and unbiased exploration of this intriguing population.

    \item Searches for dual AGNs. While our current search yielded one promising candidate radio AGN pair (J1307+0422) already known in the literature, deeper radio surveys are needed to possibly find more. The future ultra-deep, high-resolution capabilities of SKA will be perfectly suited to systematically search for dual AGNs at high redshifts, probing a key stage of galaxy merger and SMBH growth history.
    
\end{itemize}
    
While the full catalog is not flux density-limited, the FIRST-detected subsample with a well-defined detection threshold of about $1$~mJy $(6\sigma)$ can be used to investigate the radio-loud fraction as a function of redshift when compared to a similarly selected parent optical quasar sample.

\section{Conclusions}
\label{sec5:conclusion}
In this study, we constructed the RHzQCat, the largest uniformly selected sample of radio-detected quasars at $z \geq 3$ to date. The catalog was obtained by cross-matching SDSS DR16Q quasars with four major radio surveys (NVSS, FIRST, RACS, and GLEAM), and by implementing a novel multi-tiered framework to handle the large differences in angular resolution among the surveys. This strategy, combined with extensive visual inspection, enabled us to distinguish secure single component matches from complex associations, significantly improving reliability compared to simple positional matching.

The main characteristics of the RHzQCat are as follows:

\begin{itemize}
	\item Sample composition: The catalog contains 2388 sources, including 1629 robust RHRQs and 315 candidates. A total of 444 sources were rejected due to unreliable redshifts or mismatched radio associations.
	\item Resolution dependence: The reliability of the optical--radio associations closely follows the angular resolution of the radio surveys, with our adopted matching radii consistent with those applied in previous studies.
	\item Morphology: About 95\% of the identified RHRQs are compact, a higher fraction than the $\sim$90\% reported for low-redshift quasars. In addition, we identified 110 radio galaxies or RG candidates with extended or FR-like morphology, as well as one dual AGN (J1307+0422, $z\approx 3$), demonstrating the ability of our method to recover rare and valuable subgroups.
	\item Parameter space: The catalog spans $3 \leq z \leq 5.5$ and covers a rest-frame 1.4 GHz monochromatic radio luminosity range of L$_{\mathrm{1.4 GHz}} \sim 10^{25.5}$ -- $10^{29.3}$~W\,Hz$^{-1}$, highlighting the vigorous AGN activity at $z \sim$ 3 -- 4. 
	\item Sample size: Compared to previous works, the RHzQCat increases the number of known RHRQs at $z \geq 3$ by nearly an order of magnitude, forming the largest and least-biased sample to date in this redshift range.
\end{itemize}

Our results demonstrate that the RHzQCat provides a robust statistical foundation for exploring AGN evolution in the early Universe. This catalog reveals a predominantly compact HRQ population, along with a subset of extended and peculiar sources that trace powerful jet activity at ``cosmic morning''. Looking ahead, the RHzQCat offers an essential resource for high-resolution multi-wavelength follow-up observations, and serves as a benchmark for the next generation of wide-field surveys with facilities such as the Square Kilometer Array.

\begin{table}[H] 
\caption{The number of sources from different subgroups.\label{tabss:subgroup}}
\begin{adjustwidth}{-\extralength}{0cm}
\begin{tabularx}{\fulllength}{clccccccc}
\toprule
\multirow[m]{2}{*}{\textbf{match\_tier\_code \textsuperscript{1}}} & \multirow[m]{2}{*}{\textbf{Radio Surveys}} & \multicolumn{3}{C}{\textbf{Tier 1}}    & \multicolumn{3}{C}{\textbf{Tier 2/Tier 3}}   & \textbf{Total} \\
 &  & \textbf{Good} & \textbf{Cand} & \textbf{Rejected} & \textbf{Good} & \textbf{Cand} & \textbf{Rejected} &  \\
 \midrule
1000 & FIRST only & 296 & 15 & 0 & 0 & 0 & 0 & 311 \\
0100 & NVSS only & 48 & 22 & 11 & 28 & 40 & 113 & 262 \\
1100 & FIRST+NVSS & 267 & 30 & 0 & 183 & 55 & 0 & 535 \\
0010 & RACS only & 25 & 12 & 4 & 7 & 5 & 9 & 62 \\
1010 & FIRST+RACS & 65 & 5 & 0 & 18 & 2 & 0 & 90 \\
0110 & NVSS+RACS & 57 & 8 & 0 & 40 & 14 & 4 & 123 \\
1110 & FRIST+NVSS+RACS & 253 & 22 & 0 & 251 & 31 & 0 & 557 \\
0001 & GLEAM only & 0 & 0 & 0 & 0 & 9 & 214 & 223 \\
1001 & FIRST+GLEAM & 0 & 0 & 0 & 0 & 8 & 0 & 8 \\
0101 & NVSS+GLEAM & 0 & 0 & 0 & 0 & 2 & 7 & 9 \\
1101 & FRIST+NVSS+GLEAM & 0 & 0 & 0 & 2 & 5 & 0 & 7 \\
0011 & RACS+GLEAM & 0 & 0 & 0 & 0 & 1 & 0 & 1 \\
1011 & FIRST+RACS+GLEAM & 0 & 0 & 0 & 0 & 1 & 0 & 1 \\
0111 & NVSS+RACS+GLEAM & 0 & 0 & 0 & 18 & 1 & 4 & 23 \\
1111 & FIRST+NVSS+RACS+GLEAM & 0 & 0 & 0 & 71 & 27 & 0 & 98 \\
\midrule
 & Total & 1011 & 114 & 15 & 618 & 201 & 351 & 2310 \\
\bottomrule

\end{tabularx}
\end{adjustwidth}
\noindent{\footnotesize{\textsuperscript{1} See description in Table~\ref{tabs1:class1}. Note: The 78 \textit{rejected A} 
 sources are not included in this table.}} 
\end{table}

\vspace{-14pt} 
\begin{table}[H] 

\caption{Source
 groups with more than one high-$z$ quasar in the compiled catalog. \label{tabs3:dual_candidates}}
 \footnotesize
\begin{adjustwidth}{-\extralength}{0cm}
\begin{tabularx}{\fulllength}{m{3cm}<{\centering}CCCCCCC}
\toprule
\textbf{SDSS Name} & \textbf{Separation}	& \textbf{Redshift} & \textbf{Tier\boldmath{$_\mathrm{FIRST}$}} & \textbf{Tier\boldmath{$_\mathrm{NVSS}$}} & \textbf{Tier\boldmath{$_\mathrm{GLEAM}$}} & \textbf{Tier\boldmath{$_\mathrm{RACS}$}} & \textbf{Category} \\
\midrule
004905.72 $-$ 003051.2 & \multirow{2}{*}{$43.2^{\prime\prime}$ (326 kpc)} & 3.231  &  & t1 & t3 & t1 & good B\\
004905.94 $-$ 003134.3 &  & 4.165  &  &  & t3 &  & rejected B\\
\midrule
011905.36 $+$ 020512.7 & \multirow{2}{*}{$55.3^{\prime\prime}$ (426 kpc)} & 3.010  &  &  & t3 &  & rejected B\\
011907.46 $+$ 020558.2 &  & 2.970  &  &  & t3 &  & rejected B\\
\midrule
022412.54 $-$ 052730.8 & \multirow{2}{*}{$14.3^{\prime\prime}$ (101 kpc)} & 3.822  &  &  & t3 &  & rejected A\\
022413.41 $-$ 052724.8 &  & 3.779  &  &  & t3 &  &rejected B\\
\midrule
130756.18 $+$ 042215.4 & \multirow{2}{*}{$8.2^{\prime\prime}$ (63 kpc)} & 3.030  &  & t2 &  & & good \\
130756.73 $+$ 042215.5 &  & 3.012  & t1 & t2 &  & t1 & good\\
\midrule
171932.93 $+$ 291929.7 & \multirow{2}{*}{$106^{\prime\prime}$ (795 kpc)} & 3.294  &  &  &  t3& &rejected B \\
171937.86 $+$ 291805.0 &  & 3.067  &  &  & t3 &  & rejected B\\
\bottomrule
\end{tabularx}
\end{adjustwidth}
\noindent{\footnotesize{Note: Column 1---source name from DR16Q; Column 2---separation between the two optical sources in the group; Column 3---source redshift from the catalog; Columns 4$\sim$7---source matching tiers derived from our compiled catalog corresponding to radio catalogs FIRST, NVSS, GLEAM, and RACS-DR1, respectively.}}
\end{table}


\vspace{6pt} 


\authorcontributions{Conceptualization, Y.Z., T.A., and S.F.; methodology, Y.Z., R.L., and X.J.; software, Y.Z.; validation, Y.Z., R.L., K.P., and Q.W.; formal analysis, Y.Z.; investigation, Y.Z. and  K.P.; resources, Y.Z., R.L., K.P., and S.L.; data curation, Y.Z. and R.L.; writing---original draft preparation, Y.Z. and K.P.; writing---review and editing, Y.Z., R.L., K.P., S.F., T.A., X.J., Q.W., and S.L.; visualization, Y.Z.; supervision, T.A. and S.F.; funding acquisition, T.A. and S.L. All authors have read and agreed to the published version of the manuscript.}

\funding{This work is supported by the National SKA Program of China (grant no. 2022SKA0120102) and
the Strategic Priority Research Program of the Chinese Academy of Sciences, Grant No. XDA0350205.
Y.Z. is also supported by
the China Scholarship Council (No. 202104910165) and
the Shanghai Sailing Program under grant number 22YF1456100.
S.F. acknowledges funding received from the Hungarian National Research, Development and Innovation Office (NKFIH) excellence grant TKP2021-NKTA-64.
T.A. acknowledge the support from the Xinjiang Tianchi Talent Program.
L.S. is supported by the National
Natural Science Foundation of China (NSFC) through grants 12173069, the
Strategic Priority Research Program of the Chinese Academy of Sciences,
Grant No.XDA0350205, the Youth Innovation Promotion Association CAS
with Certificate Number 2022259, the Talent Plan of Shanghai Branch, Chinese
Academy of Sciences with No. CASSHB-QNPD-2023-016.}

\dataavailability{The data used in this paper were obtained from publicly available catalogs from online databases. The code and the compiled catalog were created by the authors under the MIT License. The overlay images of all sources in our catalog are publicly available via the zenodo archive (\url{https://doi.org/10.5281/zenodo.17541810}, accessed on 1 November 2025.). }

\acknowledgments{This work used resources of China SKA Regional Center prototype funded by the Ministry of Science and Technology of the People’s Republic of China and the Chinese Academy of Sciences.
Y.Z. thanks the warm host and the helpful comments from Ivy Wong in CSIRO Space\&Astronomy, Australia. 
Y.Z. thanks Fabio Luchsinger for his inspiring tutorial talk on STILTS in CSIRO's co-learnium series.}

\conflictsofinterest{The authors declare no conflicts of interest.}



\begin{adjustwidth}{-\extralength}{0cm}

\printendnotes[custom]

\reftitle{References}

\newcommand*\aap{Astron. Astrophys.}
\let\astap=\aap
\newcommand*\aapr{Astron. Astrophys. Rev.}
\newcommand*\aj{Astron. J.}
\newcommand*\ao{Appl.~Opt.}
\let\applopt\ao
\newcommand*\apj{ Astrophys. J.}
\newcommand*\apjl{ Astrophys. J. Lett.}
\let\apjlett\apjl
\newcommand*\apjs{ Astrophys. J. Suppl. Ser.}
\let\apjsupp\apjs
\newcommand*\araa{Annu. Rev. Astron. Astrophys.}
\newcommand*\mnras{Mon. Not. R. Astron. Soc. Lett.}
\newcommand*\nat{Nature}
\newcommand*\pasa{Publ. Astron. Soc. Aust.}

\newcommand*\pasp{Publ. Astron. Soc. Pac.}

\PublishersNote{}
\end{adjustwidth}

\begin{thebibliography}{999}

\bibitem[{Fabian} et~al.(2014){Fabian}, {Walker}, {Celotti}, {Ghisellini},
  {Mocz}, {Blundell}, and {McMahon}]{2014MNRAS.442L..81F}
{Fabian}, A.C.; {Walker}, S.A.; {Celotti}, A.; {Ghisellini}, G.; {Mocz}, P.;
  {Blundell}, K.M.; {McMahon}, R.G.
\newblock {Do high-redshift quasars have powerful jets?}
\newblock {\em Mon. Not. R. Astron. Soc. Lett.} {\bf 2014}, {\em 442}, L81--L84.
\newblock {\url{https://doi.org/10.1093/mnrasl/slu065}}.

\bibitem[{McQuinn}(2016)]{2016ARA&A..54..313M}
{McQuinn}, M.
\newblock {The Evolution of the Intergalactic Medium}.
\newblock {\em  Annu. Rev. Astron. Astrophys.} {\bf 2016}, {\em 54}, 313--362. 
\newblock {\url{https://doi.org/10.1146/annurev-astro-082214-122355}}.

\bibitem[{Blandford} et~al.(2019){Blandford}, {Meier}, and
  {Readhead}]{2019ARA&A..57..467B}
{Blandford}, R.; {Meier}, D.; {Readhead}, A.
\newblock {Relativistic Jets from Active Galactic Nuclei}.
\newblock {\em Annu. Rev. Astron. Astrophys.} {\bf 2019}, {\em 57},~467--509. 
\newblock {\url{https://doi.org/10.1146/annurev-astro-081817-051948}}.

\bibitem[{An} et~al.(2020){An}, {Mohan}, {Zhang}, {Frey}, {Yang},
  {Gab{\'a}nyi}, {Gurvits}, {Paragi}, {Perger}, and
  {Zheng}]{2020NatCo..11..143A}
{An}, T.; {Mohan}, P.; {Zhang}, Y.; {Frey}, S.; {Yang}, J.; {Gab{\'a}nyi},
  K.{\'E}.; {Gurvits}, L.I.; {Paragi}, Z.; {Perger}, K.; {Zheng}, Z.
\newblock {Evolving parsec-scale radio structure in the most distant blazar
  known}.
\newblock {\em Nat. Commun.} {\bf 2020}, {\em 11},~143. 
\newblock {\url{https://doi.org/10.1038/s41467-019-14093-2}}.

\bibitem[{Keller} et~al.(2024){Keller}, {Thyagarajan}, {Kumar}, {Kanekar}, and
  {Bernardi}]{2024MNRAS.528.5692K}
{Keller}, P.M.; {Thyagarajan}, N.; {Kumar}, A.; {Kanekar}, N.; {Bernardi}, G.
\newblock {The radio-loud fraction of quasars at z > 6}.
\newblock {\em Mon. Not. R. Astron. Soc.} {\bf 2024}, {\em 528}, 5692--5702. 
\newblock {\url{https://doi.org/10.1093/mnras/stae418}}.

\bibitem[{Fan} et~al.(2006){Fan}, {Strauss}, {Becker}, {White}, {Gunn},
  {Knapp}, {Richards}, {Schneider}, {Brinkmann}, and
  {Fukugita}]{2006AJ....132..117F}
{Fan}, X.; {Strauss}, M.A.; {Becker}, R.H.; {White}, R.L.; {Gunn}, J.E.;
  {Knapp}, G.R.; {Richards}, G.T.; {Schneider}, D.P.; {Brinkmann}, J.;
  {Fukugita}, M.
\newblock {Constraining the Evolution of the Ionizing Background and the Epoch
  of Reionization with z\raisebox{-0.5ex}\textasciitilde6 Quasars. II. A Sample
  of \mbox{19 Quasars}.}
\newblock {\em  Astron. J.} {\bf 2006}, {\em 132},~117--136. 
\newblock {\url{https://doi.org/10.1086/504836}}.

\bibitem[{Stoughton} et~al.(2002){Stoughton}, {Lupton}, {Bernardi}, {Blanton},
  {Burles}, {Castander}, {Connolly}, {Eisenstein}, {Frieman}, {Hennessy},
  {Hindsley}, {Ivezi{\'c}}, {Kent}, {Kunszt}, {Lee}, {Meiksin}, {Munn},
  {Newberg}, {Nichol}, {Nicinski}, {Pier}, {Richards}, {Richmond}, {Schlegel},
  {Smith}, {Strauss}, {SubbaRao}, {Szalay}, {Thakar}, {Tucker}, {Vanden Berk},
  {Yanny}, {Adelman}, {Anderson}, {Anderson}, {Annis}, {Bahcall}, {Bakken},
  {Bartelmann}, {Bastian}, {Bauer}, {Berman}, {B{\"o}hringer}, {Boroski},
  {Bracker}, {Briegel}, {Briggs}, {Brinkmann}, {Brunner}, {Carey}, {Carr},
  {Chen}, {Christian}, {Colestock}, {Crocker}, {Csabai}, {Czarapata},
  {Dalcanton}, {Davidsen}, {Davis}, {Dehnen}, {Dodelson}, {Doi}, {Dombeck},
  {Donahue}, {Ellman}, {Elms}, {Evans}, {Eyer}, {Fan}, {Federwitz}, {Friedman},
  {Fukugita}, {Gal}, {Gillespie}, {Glazebrook}, {Gray}, {Grebel}, {Greenawalt},
  {Greene}, {Gunn}, {de Haas}, {Haiman}, {Haldeman}, {Hall}, {Hamabe},
  {Hansen}, {Harris}, {Harris}, {Harvanek}, {Hawley}, {Hayes}, {Heckman},
  {Helmi}, {Henden}, {Hogan}, {Hogg}, {Holmgren}, {Holtzman}, {Huang}, {Hull},
  {Ichikawa}, {Ichikawa}, {Johnston}, {Kauffmann}, {Kim}, {Kimball}, {Kinney},
  {Klaene}, {Kleinman}, {Klypin}, {Knapp}, {Korienek}, {Krolik}, {Kron},
  {Krzesi{\'n}ski}, {Lamb}, {Leger}, {Limmongkol}, {Lindenmeyer}, {Long},
  {Loomis}, {Loveday}, {MacKinnon}, {Mannery}, {Mantsch}, {Margon}, {McGehee},
  {McKay}, {McLean}, {Menou}, {Merelli}, {Mo}, {Monet}, {Nakamura},
  {Narayanan}, {Nash}, {Neilsen}, {Newman}, {Nitta}, {Odenkirchen}, {Okada},
  {Okamura}, {Ostriker}, {Owen}, {Pauls}, {Peoples}, {Peterson}, {Petravick},
  {Pope}, {Pordes}, {Postman}, {Prosapio}, {Quinn}, {Rechenmacher}, {Rivetta},
  {Rix}, {Rockosi}, {Rosner}, {Ruthmansdorfer}, {Sandford}, {Schneider},
  {Scranton}, {Sekiguchi}, {Sergey}, {Sheth}, {Shimasaku}, {Smee}, {Snedden},
  {Stebbins}, {Stubbs}, {Szapudi}, {Szkody}, {Szokoly}, {Tabachnik},
  {Tsvetanov}, {Uomoto}, {Vogeley}, {Voges}, {Waddell}, {Walterbos}, {Wang},
  {Watanabe}, {Weinberg}, {White}, {White}, {Wilhite}, {Wolfe}, {Yasuda},
  {York}, {Zehavi}, and {Zheng}]{2002AJ....123..485S}
{Stoughton}, C.; {Lupton}, R.H.; {Bernardi}, M.; {Blanton}, M.R.; {Burles}, S.;
  {Castander}, F.J.; {Connolly}, A.J.; {Eisenstein}, D.J.; {Frieman}, J.A.;
  {Hennessy}, G.S.;  et~al.
\newblock {Sloan Digital Sky Survey: Early Data Release}.
\newblock {\em Astron. J.} {\bf 2002}, {\em 123},~485--548.
\newblock {\url{https://doi.org/10.1086/324741}}.

\bibitem[{Kaiser} et~al.(2002){Kaiser}, {Aussel}, {Burke}, {Boesgaard},
  {Chambers}, {Chun}, {Heasley}, {Hodapp}, {Hunt}, {Jedicke}, {Jewitt},
  {Kudritzki}, {Luppino}, {Maberry}, {Magnier}, {Monet}, {Onaka}, {Pickles},
  {Rhoads}, {Simon}, {Szalay}, {Szapudi}, {Tholen}, {Tonry}, {Waterson}, and
  {Wick}]{2002SPIE.4836..154K}
{Kaiser}, N.; {Aussel}, H.; {Burke}, B.E.; {Boesgaard}, H.; {Chambers}, K.;
  {Chun}, M.R.; {Heasley}, J.N.; {Hodapp}, K.W.; {Hunt}, B.; {Jedicke}, R.;
  et~al.
\newblock {Pan-STARRS: A Large Synoptic Survey Telescope Array}.
\newblock In Proceedings of the Survey and Other Telescope Technologies and
  Discoveries, Waikoloa, HI, USA, 27--28 August 2002;
 {Tyson}, J.A., {Wolff}, S., Eds.; Society
  of Photo-Optical Instrumentation Engineers (SPIE) Conference Series: Bellingham, WA, USA,
 2002; Volume 4836, pp. 154--164.
\newblock {\url{https://doi.org/10.1117/12.457365}}.

\bibitem[{Kaiser} et~al.(2010){Kaiser}, {Burgett}, {Chambers}, {Denneau},
  {Heasley}, {Jedicke}, {Magnier}, {Morgan}, {Onaka}, and
  {Tonry}]{2010SPIE.7733E..0EK}
{Kaiser}, N.; {Burgett}, W.; {Chambers}, K.; {Denneau}, L.; {Heasley}, J.;
  {Jedicke}, R.; {Magnier}, E.; {Morgan}, J.; {Onaka}, P.; {Tonry}, J.
\newblock {The Pan-STARRS wide-field optical/NIR imaging survey}.
\newblock In Proceedings of the Ground-based and Airborne Telescopes III, San Diego, CA, USA, 17 June--2 July 2010;
  {Stepp}, L.M., {Gilmozzi}, R., {Hall}, H.J., Eds.;   Society of Photo-Optical Instrumentation Engineers (SPIE) Conference Series: Bellingham, WA, USA,
   2010; Volume 7733,  p. 77330E.
\newblock {\url{https://doi.org/10.1117/12.859188}}.

\bibitem[{Ross} and {Cross}(2020)]{2020MNRAS.494..789R}
{Ross}, N.P.; {Cross}, N.J.G.
\newblock {The near and mid-infrared photometric properties of known redshift z
  {\ensuremath{\geq}} 5 quasars}.
\newblock {\em \mnras} {\bf 2020}, {\em 494},~789--803. 
\newblock {\url{https://doi.org/10.1093/mnras/staa544}}.

\bibitem[{Onorato} et~al.(2025){Onorato}, {Hennawi}, {Schindler}, {Yang},
  {Wang}, {Barth}, {Ba{\~n}ados}, {Eilers}, {Bosman}, {Davies}, {Venemans},
  {Mazzucchelli}, {Belladitta}, {Vito}, {Farina}, {Andika}, {Fan}, {Walter},
  {Decarli}, {Onoue}, and {Nanni}]{2025MNRAS.540.1308O}
{Onorato}, S.; {Hennawi}, J.F.; {Schindler}, J.T.; {Yang}, J.; {Wang}, F.;
  {Barth}, A.J.; {Ba{\~n}ados}, E.; {Eilers}, A.C.; {Bosman}, S.E.I.; {Davies},
  F.B.;  et~al.
\newblock {Optical and near-infrared spectroscopy of quasars at z > 6.5: Public
  data release and composite spectrum}.
\newblock {\em \mnras} {\bf 2025}, {\em 540},~1308--1328.
\newblock {\url{https://doi.org/10.1093/mnras/staf787}}.

\bibitem[{Willott} et~al.(2010){Willott}, {Delorme}, {Reyl{\'e}}, {Albert},
  {Bergeron}, {Crampton}, {Delfosse}, {Forveille}, {Hutchings}, {McLure},
  {Omont}, and {Schade}]{2010AJ....139..906W}
{Willott}, C.J.; {Delorme}, P.; {Reyl{\'e}}, C.; {Albert}, L.; {Bergeron}, J.;
  {Crampton}, D.; {Delfosse}, X.; {Forveille}, T.; {Hutchings}, J.B.; {McLure},
  R.J.;  et~al.
\newblock {The Canada-France High-z Quasar Survey: Nine New Quasars and the
  Luminosity Function at Redshift 6}.
\newblock {\em \aj} {\bf 2010}, {\em 139},~906--918. 
\newblock {\url{https://doi.org/10.1088/0004-6256/139/3/906}}.

\bibitem[{Yang} et~al.(2017){Yang}, {Fan}, {Wu}, {Wang}, {Bian}, {Yang},
  {McGreer}, {Yi}, {Jiang}, {Green}, {Yue}, {Wang}, {Li}, {Ding}, {Dye}, and
  {Lawrence}]{2017AJ....153..184Y}
{Yang}, J.; {Fan}, X.; {Wu}, X.B.; {Wang}, F.; {Bian}, F.; {Yang}, Q.;
  {McGreer}, I.D.; {Yi}, W.; {Jiang}, L.; {Green}, R.;  et~al.
\newblock {Discovery of 16 New \mbox{z {\ensuremath{\sim}} 5.5} Quasars: Filling in
  the Redshift Gap of Quasar Color Selection}.
\newblock {\em \aj} {\bf 2017}, {\em 153},~184. 
\newblock {\url{https://doi.org/10.3847/1538-3881/aa6577}}.

\bibitem[{Yang} et~al.(2023){Yang}, {Fan}, {Gupta}, {Myers},
  {Palanque-Delabrouille}, {Wang}, {Y{\`e}che}, {Aguilar}, {Ahlen},
  {Alexander}, {Brooks}, {Dawson}, {de la Macorra}, {Dey}, {Dhungana},
  {Fanning}, {Font-Ribera}, {Gontcho}, {Guy}, {Honscheid}, {Juneau}, {Kisner},
  {Kremin}, {Le Guillou}, {Levi}, {Magneville}, {Martini}, {Meisner}, {Miquel},
  {Moustakas}, {Nie}, {Percival}, {Poppett}, {Prada}, {Schlafly}, {Tarl{\'e}},
  {Vargas Magana}, {Weaver}, {Wechsler}, {Zhou}, {Zhou}, and
  {Zou}]{2023ApJS..269...27Y}
{Yang}, J.; {Fan}, X.; {Gupta}, A.; {Myers}, A.D.; {Palanque-Delabrouille}, N.;
  {Wang}, F.; {Y{\`e}che}, C.; {Aguilar}, J.N.; {Ahlen}, S.; {Alexander}, D.M.;
   et~al.
\newblock {DESI z {\ensuremath{\gtrsim}} 5 Quasar Survey. I. A First Sample of
  400 New Quasars at z 4.7--6.6}.
\newblock {\em  Astrophys. J. Suppl. Ser.} {\bf 2023}, {\em 269},~27. 
\newblock {\url{https://doi.org/10.3847/1538-4365/acf99b}}.

\bibitem[{Spingola} et~al.(2020){Spingola}, {Dallacasa}, {Belladitta},
  {Caccianiga}, {Giroletti}, {Moretti}, and {Orienti}]{2020A&A...643L..12S}
{Spingola}, C.; {Dallacasa}, D.; {Belladitta}, S.; {Caccianiga}, A.;
  {Giroletti}, M.; {Moretti}, A.; {Orienti}, M.
\newblock {Parsec-scale properties of the radio brightest jetted AGN at z > 6}.
\newblock {\em \aap} {\bf 2020}, {\em 643},~L12. 
\newblock {\url{https://doi.org/10.1051/0004-6361/202039458}}.

\bibitem[{Zhang} et~al.(2022){Zhang}, {An}, {Frey}, {Gab{\'a}nyi}, and
  {Sotnikova}]{2022ApJ...937...19Z}
{Zhang}, Y.; {An}, T.; {Frey}, S.; {Gab{\'a}nyi}, K.{\'E}.; {Sotnikova}, Y.
\newblock {Radio Jet Proper-motion Analysis of Nine Distant Quasars above
  Redshift 3.5}.
\newblock {\em \apj} {\bf 2022}, {\em 937},~19. 
\newblock {\url{https://doi.org/10.3847/1538-4357/ac87f8}}.

\bibitem[{Ighina} et~al.(2025){Ighina}, {Caccianiga}, {Moretti}, {Broderick},
  {Leung}, {Rigamonti}, {Seymour}, {Afonso}, {Connor}, {Vignali}, {Wang}, {An},
  {Arsioli}, {Bisogni}, {Dallacasa}, {Della Ceca}, {Liu},
  {L{\'o}pez-S{\'a}nchez}, {Matute}, {Reynolds}, {Rossi}, {Spingola},
  {Severgnini}, and {Tavecchio}]{2025A&A...698A.158I}
{Ighina}, L.; {Caccianiga}, A.; {Moretti}, A.; {Broderick}, J.W.; {Leung},
  J.K.; {Rigamonti}, F.; {Seymour}, N.; {Afonso}, J.; {Connor}, T.; {Vignali},
  C.;  et~al.
\newblock {High-z radio quasars in RACS: I. Selection, identification, and
  multi-wavelength properties}.
\newblock {\em \aap} {\bf 2025}, {\em 698},~A158. 
\newblock {\url{https://doi.org/10.1051/0004-6361/202453650}}.

\bibitem[{Ba{\~n}ados} et~al.(2025){Ba{\~n}ados}, {Momjian}, {Connor},
  {Belladitta}, {Decarli}, {Mazzucchelli}, {Venemans}, {Walter}, {Wang}, {Xie},
  {Barth}, {Eilers}, {Fan}, {Khusanova}, {Schindler}, {Stern}, {Yang},
  {Andika}, {Carilli}, {Farina}, {Fabian}, {Hennawi}, {Pensabene}, and
  {Rojas-Ruiz}]{2025NatAs...9..293B}
{Ba{\~n}ados}, E.; {Momjian}, E.; {Connor}, T.; {Belladitta}, S.; {Decarli},
  R.; {Mazzucchelli}, C.; {Venemans}, B.P.; {Walter}, F.; {Wang}, F.; {Xie},
  Z.L.;  et~al.
\newblock {A blazar in the epoch of reionization}.
\newblock {\em Nat. Astron.} {\bf 2025}, {\em 9},~293--301. 
\newblock {\url{https://doi.org/10.1038/s41550-024-02431-4}}.

\bibitem[{Kellermann}(1993)]{1993Natur.361..134K}
{Kellermann}, K.I.
\newblock {The cosmological deceleration parameter estimated from the
  angular-size/redshift relation for compact radio sources}.
\newblock {\em \nat} {\bf 1993}, {\em 361},~134--136.
\newblock {\url{https://doi.org/10.1038/361134a0}}.

\bibitem[{Ivezi{\'c}} et~al.(2002){Ivezi{\'c}}, {Menou}, {Knapp}, {Strauss},
  {Lupton}, {Vanden Berk}, {Richards}, {Tremonti}, {Weinstein}, {Anderson},
  {Bahcall}, {Becker}, {Bernardi}, {Blanton}, {Eisenstein}, {Fan},
  {Finkbeiner}, {Finlator}, {Frieman}, {Gunn}, {Hall}, {Kim}, {Kinkhabwala},
  {Narayanan}, {Rockosi}, {Schlegel}, {Schneider}, {Strateva}, {SubbaRao},
  {Thakar}, {Voges}, {White}, {Yanny}, {Brinkmann}, {Doi}, {Fukugita},
  {Hennessy}, {Munn}, {Nichol}, and {York}]{2002AJ....124.2364I}
{Ivezi{\'c}}, {\v{Z}}.; {Menou}, K.; {Knapp}, G.R.; {Strauss}, M.A.; {Lupton},
  R.H.; {Vanden Berk}, D.E.; {Richards}, G.T.; {Tremonti}, C.; {Weinstein},
  M.A.; {Anderson}, S.;  et~al.
\newblock {Optical and Radio Properties of Extragalactic Sources Observed by
  the FIRST Survey and the Sloan Digital Sky Survey}.
\newblock {\em \aj} {\bf 2002}, {\em 124},~2364--2400. 
\newblock {\url{https://doi.org/10.1086/344069}}.

\bibitem[{Kellermann} et~al.(2016){Kellermann}, {Condon}, {Kimball}, {Perley},
  and {Ivezi{\'c}}]{2016ApJ...831..168K}
{Kellermann}, K.I.; {Condon}, J.J.; {Kimball}, A.E.; {Perley}, R.A.;
  {Ivezi{\'c}}, {\v{Z}}.
\newblock {Radio-loud and Radio-quiet QSOs}.
\newblock {\em \apj} {\bf 2016}, {\em 831},~168. 
\newblock {\url{https://doi.org/10.3847/0004-637X/831/2/168}}.

\bibitem[{Arsenov} et~al.(2025){Arsenov}, {Frey}, {Kov{\'a}cs}, and
  {Slavcheva-Mihova}]{2025ApJS..280...23A}
{Arsenov}, N.; {Frey}, S.; {Kov{\'a}cs}, A.; {Slavcheva-Mihova}, L.
\newblock {Radio-loudness Statistics of Quasars from Quaia-VLASS}.
\newblock {\em \apjs} {\bf 2025}, {\em 280},~23. 
\newblock {\url{https://doi.org/10.3847/1538-4365/adf056}}.

\bibitem[{Paliya} et~al.(2020){Paliya}, {Ajello}, {Cao}, {Giroletti}, {Kaur},
  {Madejski}, {Lott}, and {Hartmann}]{2020ApJ...897..177P}
{Paliya}, V.S.; {Ajello}, M.; {Cao}, H.M.; {Giroletti}, M.; {Kaur}, A.;
  {Madejski}, G.; {Lott}, B.; {Hartmann}, D.
\newblock {Blazars at the Cosmic Dawn}.
\newblock {\em \apj} {\bf 2020}, {\em 897},~177. 
\newblock {\url{https://doi.org/10.3847/1538-4357/ab9c1a}}.

\bibitem[{Trakhtenbrot} et~al.(2017){Trakhtenbrot}, {Volonteri}, and
  {Natarajan}]{2017ApJ...836L...1T}
{Trakhtenbrot}, B.; {Volonteri}, M.; {Natarajan}, P.
\newblock {On the Accretion Rates and Radiative Efficiencies of the
  Highest-redshift Quasars}.
\newblock {\em \apjl} {\bf 2017}, {\em 836},~L1. 
\newblock {\url{https://doi.org/10.3847/2041-8213/836/1/L1}}.

\bibitem[{Perger} et~al.(2017){Perger}, {Frey}, {Gab{\'a}nyi}, and
  {T{\'o}th}]{2017FrASS...4....9P}
{Perger}, K.; {Frey}, S.; {Gab{\'a}nyi}, K.{\'E}.; {T{\'o}th}, L.V.
\newblock {A catalogue of active galactic nuclei from the first 1.5 Gyr of the
  Universe}.
\newblock {\em Front. Astron. Space Sci.} {\bf 2017}, {\em  4},~9.
\newblock {\url{https://doi.org/10.3389/fspas.2017.00009}}.

\bibitem[{Perger} et~al.(2024){Perger}, {Frey}, and
  {Gab{\'a}nyi}]{2024MNRAS.527.3436P}
{Perger}, K.; {Frey}, S.; {Gab{\'a}nyi}, K.{\'E}.
\newblock {Sub-mJy radio emission from high-redshift active galactic nuclei in
  the footprint of the VLA Sky Survey}.
\newblock {\em \mnras} {\bf 2024}, {\em 527},~3436--3444.
\newblock {\url{https://doi.org/10.1093/mnras/stad3411}}.

\bibitem[{Krezinger} et~al.(2022){Krezinger}, {Perger}, {Gab{\'a}nyi}, {Frey},
  {Gurvits}, {Paragi}, {An}, {Zhang}, {Cao}, and
  {Sbarrato}]{2022ApJS..260...49K}
{Krezinger}, M.; {Perger}, K.; {Gab{\'a}nyi}, K.{\'E}.; {Frey}, S.; {Gurvits},
  L.I.; {Paragi}, Z.; {An}, T.; {Zhang}, Y.; {Cao}, H.; {Sbarrato}, T.
\newblock {Radio-loud Quasars above Redshift 4: Very Long Baseline
  Interferometry (VLBI) Imaging of an Extended Sample}.
\newblock {\em \apjs} {\bf 2022}, {\em 260},~49. 
\newblock {\url{https://doi.org/10.3847/1538-4365/ac63b8}}.

\bibitem[{Capetti} and {Balmaverde}(2024)]{2024A&A...689A.174C}
{Capetti}, A.; {Balmaverde}, B.
\newblock {The radio properties of z > 3.5 quasars: Are most high-redshift
  radio-loud active galactic nuclei obscured?}
\newblock {\em \aap} {\bf 2024}, {\em 689},~A174. 
\newblock {\url{https://doi.org/10.1051/0004-6361/202449676}}.

\bibitem[{Becker} et~al.(1995){Becker}, {White}, and
  {Helfand}]{1995ApJ...450..559B}
{Becker}, R.H.; {White}, R.L.; {Helfand}, D.J.
\newblock {The FIRST Survey: Faint Images of the Radio Sky at Twenty
  Centimeters}.
\newblock {\em \apj} {\bf 1995}, {\em 450},~559.
\newblock {\url{https://doi.org/10.1086/176166}}.

\bibitem[{Lacy} et~al.(2020){Lacy}, {Baum}, {Chandler}, {Chatterjee}, {Clarke},
  {Deustua}, {English}, {Farnes}, {Gaensler}, {Gugliucci}, {Hallinan}, {Kent},
  {Kimball}, {Law}, {Lazio}, {Marvil}, {Mao}, {Medlin}, {Mooley}, {Murphy},
  {Myers}, {Osten}, {Richards}, {Rosolowsky}, {Rudnick}, {Schinzel},
  {Sivakoff}, {Sjouwerman}, {Taylor}, {White}, {Wrobel}, {Andernach},
  {Beasley}, {Berger}, {Bhatnager}, {Birkinshaw}, {Bower}, {Brandt}, {Brown},
  {Burke-Spolaor}, {Butler}, {Comerford}, {Demorest}, {Fu}, {Giacintucci},
  {Golap}, {G{\"u}th}, {Hales}, {Hiriart}, {Hodge}, {Horesh}, {Ivezi{\'c}},
  {Jarvis}, {Kamble}, {Kassim}, {Liu}, {Loinard}, {Lyons}, {Masters}, {Mezcua},
  {Moellenbrock}, {Mroczkowski}, {Nyland}, {O'Dea}, {O'Sullivan}, {Peters},
  {Radford}, {Rao}, {Robnett}, {Salcido}, {Shen}, {Sobotka}, {Witz}, {Vaccari},
  {van Weeren}, {Vargas}, {Williams}, and {Yoon}]{2020PASP..132c5001L}
{Lacy}, M.; {Baum}, S.A.; {Chandler}, C.J.; {Chatterjee}, S.; {Clarke}, T.E.;
  {Deustua}, S.; {English}, J.; {Farnes}, J.; {Gaensler}, B.M.; {Gugliucci},
  N.;  et~al.
\newblock {The Karl G. Jansky Very Large Array Sky Survey (VLASS). Science Case
  and Survey Design}.
\newblock {\em \pasp} {\bf 2020}, {\em 132},~035001. 
\newblock {\url{https://doi.org/10.1088/1538-3873/ab63eb}}.

\bibitem[{Hale} et~al.(2021){Hale}, {McConnell}, {Thomson}, {Lenc}, {Heald},
  {Hotan}, {Leung}, {Moss}, {Murphy}, {Pritchard}, {Sadler}, {Stewart}, and
  {Whiting}]{2021PASA...38...58H}
{Hale}, C.L.; {McConnell}, D.; {Thomson}, A.J.M.; {Lenc}, E.; {Heald}, G.H.;
  {Hotan}, A.W.; {Leung}, J.K.; {Moss}, V.A.; {Murphy}, T.; {Pritchard}, J.;
  et~al.
\newblock {The Rapid ASKAP Continuum Survey Paper II: First Stokes I Source
  Catalogue Data Release}.
\newblock {\em \pasa} {\bf 2021}, {\em 38},~e058. 
\newblock {\url{https://doi.org/10.1017/pasa.2021.47}}.

\bibitem[{Duchesne} et~al.(2024){Duchesne}, {Grundy}, {Heald}, {Lenc}, {Leung},
  {McConnell}, {Murphy}, {Pritchard}, {Rose}, {Thomson}, {Wang}, {Wang}, and
  {Whiting}]{2024PASA...41....3D}
{Duchesne}, S.W.; {Grundy}, J.A.; {Heald}, G.H.; {Lenc}, E.; {Leung}, J.K.;
  {McConnell}, D.; {Murphy}, T.; {Pritchard}, J.; {Rose}, K.; {Thomson},
  A.J.M.;  et~al.
\newblock {The Rapid ASKAP Continuum Survey V: Cataloguing the sky at 1 367.5
  MHz and the second data release of RACS-mid}.
\newblock {\em \pasa} {\bf 2024}, {\em 41},~e003. 
\newblock {\url{https://doi.org/10.1017/pasa.2023.60}}.

\bibitem[{Duchesne} et~al.(2025){Duchesne}, {Ross}, {Thomson}, {Lenc},
  {Murphy}, {Galvin}, {Hotan}, {Moss}, and {Whiting}]{2025PASA...42...38D}
{Duchesne}, S.; {Ross}, K.; {Thomson}, A.J.M.; {Lenc}, E.; {Murphy}, T.;
  {Galvin}, T.J.; {Hotan}, A.W.; {Moss}, V.A.; {Whiting}, M.T.
\newblock {The Rapid ASKAP Continuum Survey (RACS) VI: The RACS-high 1655.5 MHz
  images and catalogue.}
\newblock {\em \pasa} {\bf 2025}, {\em 42},~38. 
\newblock {\url{https://doi.org/10.1017/pasa.2025.2}}.

\bibitem[{Condon} et~al.(1998){Condon}, {Cotton}, {Greisen}, {Yin}, {Perley},
  {Taylor}, and {Broderick}]{1998AJ....115.1693C}
{Condon}, J.J.; {Cotton}, W.D.; {Greisen}, E.W.; {Yin}, Q.F.; {Perley}, R.A.;
  {Taylor}, G.B.; {Broderick}, J.J.
\newblock {The NRAO VLA Sky Survey}.
\newblock {\em \aj} {\bf 1998}, {\em 115},~1693--1716.
\newblock {\url{https://doi.org/10.1086/300337}}.

\bibitem[{Hurley-Walker} et~al.(2017){Hurley-Walker}, {Callingham}, {Hancock},
  {Franzen}, {Hindson}, {Kapi{\'n}ska}, {Morgan}, {Offringa}, {Wayth}, {Wu},
  {Zheng}, {Murphy}, {Bell}, {Dwarakanath}, {For}, {Gaensler},
  {Johnston-Hollitt}, {Lenc}, {Procopio}, {Staveley-Smith}, {Ekers}, {Bowman},
  {Briggs}, {Cappallo}, {Deshpande}, {Greenhill}, {Hazelton}, {Kaplan},
  {Lonsdale}, {McWhirter}, {Mitchell}, {Morales}, {Morgan}, {Oberoi}, {Ord},
  {Prabu}, {Shankar}, {Srivani}, {Subrahmanyan}, {Tingay}, {Webster},
  {Williams}, and {Williams}]{2017MNRAS.464.1146H}
{Hurley-Walker}, N.; {Callingham}, J.R.; {Hancock}, P.J.; {Franzen}, T.M.O.;
  {Hindson}, L.; {Kapi{\'n}ska}, A.D.; {Morgan}, J.; {Offringa}, A.R.; {Wayth},
  R.B.; {Wu}, C.;  et~al.
\newblock {GaLactic and Extragalactic All-sky Murchison Widefield Array (GLEAM)
  survey - I. A low-frequency extragalactic catalogue}.
\newblock {\em \mnras} {\bf 2017}, {\em 464},~1146--1167. 
\newblock {\url{https://doi.org/10.1093/mnras/stw2337}}.

\bibitem[{Lyke} et~al.(2020){Lyke}, {Higley}, {McLane}, {Schurhammer}, {Myers},
  {Ross}, {Dawson}, {Chabanier}, {Martini}, {Busca}, {Mas des Bourboux},
  {Salvato}, {Streblyanska}, {Zarrouk}, {Burtin}, {Anderson}, {Bautista},
  {Bizyaev}, {Brandt}, {Brinkmann}, {Brownstein}, {Comparat}, {Green}, {de la
  Macorra}, {Mu{\~n}oz Guti{\'e}rrez}, {Hou}, {Newman},
  {Palanque-Delabrouille}, {P{\^a}ris}, {Percival}, {Petitjean}, {Rich},
  {Rossi}, {Schneider}, {Smith}, {Vivek}, and {Weaver}]{2020ApJS..250....8L}
{Lyke}, B.W.; {Higley}, A.N.; {McLane}, J.N.; {Schurhammer}, D.P.; {Myers},
  A.D.; {Ross}, A.J.; {Dawson}, K.; {Chabanier}, S.; {Martini}, P.; {Busca},
  N.G.;  et~al.
\newblock {The Sloan Digital Sky Survey Quasar Catalog: Sixteenth Data
  Release}.
\newblock {\em \apjs} {\bf 2020}, {\em 250},~8. 
\newblock {\url{https://doi.org/10.3847/1538-4365/aba623}}.

\bibitem[{Ahumada} et~al.(2020){Ahumada}, {Allende Prieto}, {Almeida},
  {Anders}, {Anderson}, {Andrews}, {Anguiano}, {Arcodia}, {Armengaud},
  {Aubert}, {Avila}, {Avila-Reese}, {Badenes}, {Balland}, {Barger},
  {Barrera-Ballesteros}, {Basu}, {Bautista}, {Beaton}, {Beers}, {Benavides},
  {Bender}, {Bernardi}, {Bershady}, {Beutler}, {Bidin}, {Bird}, {Bizyaev},
  {Blanc}, {Blanton}, {Boquien}, {Borissova}, {Bovy}, {Brandt}, {Brinkmann},
  {Brownstein}, {Bundy}, {Bureau}, {Burgasser}, {Burtin}, {Cano-D{\'\i}az},
  {Capasso}, {Cappellari}, {Carrera}, {Chabanier}, {Chaplin}, {Chapman},
  {Cherinka}, {Chiappini}, {Doohyun Choi}, {Chojnowski}, {Chung}, {Clerc},
  {Coffey}, {Comerford}, {Comparat}, {da Costa}, {Cousinou}, {Covey}, {Crane},
  {Cunha}, {Ilha}, {Dai}, {Damsted}, {Darling}, {Davidson}, {Davies}, {Dawson},
  {De}, {de la Macorra}, {De Lee}, {Queiroz}, {Deconto Machado}, {de la Torre},
  {Dell'Agli}, {du Mas des Bourboux}, {Diamond-Stanic}, {Dillon}, {Donor},
  {Drory}, {Duckworth}, {Dwelly}, {Ebelke}, {Eftekharzadeh}, {Davis Eigenbrot},
  {Elsworth}, {Eracleous}, {Erfanianfar}, {Escoffier}, {Fan}, {Farr},
  {Fern{\'a}ndez-Trincado}, {Feuillet}, {Finoguenov}, {Fofie},
  {Fraser-McKelvie}, {Frinchaboy}, {Fromenteau}, {Fu}, {Galbany}, {Garcia},
  {Garc{\'\i}a-Hern{\'a}ndez}, {Garma Oehmichen}, {Ge}, {Geimba Maia},
  {Geisler}, {Gelfand}, {Goddy}, {Gonzalez-Perez}, {Grabowski}, {Green},
  {Grier}, {Guo}, {Guy}, {Harding}, {Hasselquist}, {Hawken}, {Hayes}, {Hearty},
  {Hekker}, {Hogg}, {Holtzman}, {Horta}, {Hou}, {Hsieh}, {Huber}, {Hunt}, {Ider
  Chitham}, {Imig}, {Jaber}, {Jimenez Angel}, {Johnson}, {Jones},
  {J{\"o}nsson}, {Jullo}, {Kim}, {Kinemuchi}, {Kirkpatrick}, {Kite}, {Klaene},
  {Kneib}, {Kollmeier}, {Kong}, {Kounkel}, {Krishnarao}, {Lacerna}, {Lan},
  {Lane}, {Law}, {Le Goff}, {Leung}, {Lewis}, {Li}, {Lian}, {Lin}, {Long},
  {Longa-Pe{\~n}a}, {Lundgren}, {Lyke}, {Mackereth}, {MacLeod}, {Majewski},
  {Manchado}, {Maraston}, {Martini}, {Masseron}, {Masters}, {Mathur},
  {McDermid}, {Merloni}, {Merrifield}, {M{\'e}sz{\'a}ros}, {Miglio}, {Minniti},
  {Minsley}, {Miyaji}, {Mohammad}, {Mosser}, {Mueller}, {Muna},
  {Mu{\~n}oz-Guti{\'e}rrez}, {Myers}, {Nadathur}, {Nair}, {Nandra}, {Correa do
  Nascimento}, {Nevin}, {Newman}, {Nidever}, {Nitschelm}, {Noterdaeme},
  {O'Connell}, {Olmstead}, {Oravetz}, {Oravetz}, {Osorio}, {Pace}, {Padilla},
  {Palanque-Delabrouille}, and {Palicio}]{2020ApJS..249....3A}
{Ahumada}, R.; {Allende Prieto}, C.; {Almeida}, A.; {Anders}, F.; {Anderson},
  S.F.; {Andrews}, B.H.; {Anguiano}, B.; {Arcodia}, R.; {Armengaud}, E.;
  {Aubert}, M.;  et~al.
\newblock {The 16th Data Release of the Sloan Digital Sky Surveys: First
  Release from the APOGEE-2 Southern Survey and Full Release of eBOSS Spectra}.
\newblock {\em \apjs} {\bf 2020}, {\em 249},~3. 
\newblock {\url{https://doi.org/10.3847/1538-4365/ab929e}}.

\bibitem[{Padovani}(2016)]{2016A&ARv..24...13P}
{Padovani}, P.
\newblock {The faint radio sky: Radio astronomy becomes mainstream}.
\newblock {\em \aapr} {\bf 2016}, {\em 24},~13.
\newblock {\url{https://doi.org/10.1007/s00159-016-0098-6}}.

\bibitem[{Condon}(1992)]{1992ARA&A..30..575C}
{Condon}, J.J.
\newblock {Radio emission from normal galaxies.}
\newblock {\em \araa} {\bf 1992}, {\em 30},~575--611.
\newblock {\url{https://doi.org/10.1146/annurev.aa.30.090192.003043}}.

\bibitem[{White} et~al.(1997){White}, {Becker}, {Helfand}, and
  {Gregg}]{1997ApJ...475..479W}
{White}, R.L.; {Becker}, R.H.; {Helfand}, D.J.; {Gregg}, M.D.
\newblock {A Catalog of 1.4 GHz Radio Sources from the FIRST Survey}.
\newblock {\em \apj} {\bf 1997}, {\em 475},~479--493.
\newblock {\url{https://doi.org/10.1086/303564}}.

\bibitem[{Taylor}(2006)]{2006ASPC..351..666T}
{Taylor}, M.B.
\newblock {STILTS---A Package for Command-Line Processing of Tabular Data}.
\newblock In Proceedings of the Astronomical Data Analysis Software and Systems
  XV, Madrid, Spain, 2--5 October 2005; 
 {Gabriel}, C., {Arviset}, C., {Ponz}, D., {Enrique}, S., Eds.; Astronomical Society of the Pacific Conference Series: San Francisco, CA, USA, 
 2006; Volume 351, p.  666.

\bibitem[{Taylor}(2005)]{2005ASPC..347...29T}
{Taylor}, M.B.
\newblock {TOPCAT \& STIL: Starlink Table/VOTable Processing Software}.
\newblock In Proceedings of the Astronomical Data Analysis Software and Systems
  XIV, Pasadena, CA, USA, 24--27 October 2004;   
   {Shopbell}, P., {Britton}, M., {Ebert}, R., Eds.; Astronomical Society of the Pacific Conference Series: Pasadena, CA, USA, 
 2005; Volume 347, p.~29.

\bibitem[{Ye} et~al.(2024){Ye}, {Zhang}, and {Wu}]{2024ApJS..275...19Y}
{Ye}, G.; {Zhang}, H.; {Wu}, Q.
\newblock {Machine Learning{\textendash}based Search of High-redshift Quasars}.
\newblock {\em \apjs} {\bf 2024}, {\em 275},~19. 
\newblock {\url{https://doi.org/10.3847/1538-4365/ad79ee}}.

\bibitem[{Frey} et~al.(2015){Frey}, {Paragi}, {Fogasy}, and
  {Gurvits}]{2015MNRAS.446.2921F}
{Frey}, S.; {Paragi}, Z.; {Fogasy}, J.O.; {Gurvits}, L.I.
\newblock {The first estimate of radio jet proper motion at z > 5}.
\newblock {\em \mnras} {\bf 2015}, {\em 446},~2921--2928. 
\newblock {\url{https://doi.org/10.1093/mnras/stu2294}}.

\bibitem[{Kozie{\l}-Wierzbowska} et~al.(2020){Kozie{\l}-Wierzbowska}, {Goyal},
  and {{\.Z}ywucka}]{2020ApJS..247...53K}
{Kozie{\l}-Wierzbowska}, D.; {Goyal}, A.; {{\.Z}ywucka}, N.
\newblock {Radio Sources Associated with Optical Galaxies and Having Unresolved
  or Extended Morphologies (ROGUE). I. A Catalog of SDSS Galaxies with FIRST
  Core Identifications}.
\newblock {\em \apjs} {\bf 2020}, {\em 247},~53. 
\newblock {\url{https://doi.org/10.3847/1538-4365/ab63d3}}.

\bibitem[{Fanaroff} and {Riley}(1974)]{1974MNRAS.167P..31F}
{Fanaroff}, B.L.; {Riley}, J.M.
\newblock {The morphology of extragalactic radio sources of high and low
  luminosity}.
\newblock {\em \mnras} {\bf 1974}, {\em 167},~31P--36P.
\newblock {\url{https://doi.org/10.1093/mnras/167.1.31P}}.

\bibitem[{Sadler} et~al.(1999){Sadler}, {McIntyre}, {Jackson}, and
  {Cannon}]{1999PASA...16..247S}
{Sadler}, E.M.; {McIntyre}, V.J.; {Jackson}, C.A.; {Cannon}, R.D.
\newblock {Radio Sources in the 2dF Galaxy Redshift Survey I. Radio Source
  Populations*}.
\newblock {\em \pasa} {\bf 1999}, {\em 16},~247--256. 
\newblock {\url{https://doi.org/10.1071/AS99247}}.

\bibitem[{Urry} and {Padovani}(1995)]{1995PASP..107..803U}
{Urry}, C.M.; {Padovani}, P.
\newblock {Unified Schemes for Radio-Loud Active Galactic Nuclei}.
\newblock {\em \pasp} {\bf 1995}, {\em 107},~803. 
\newblock {\url{https://doi.org/10.1086/133630}}.

\bibitem[{O'Dea} and {Saikia}(2021)]{2021A&ARv..29....3O}
{O'Dea}, C.P.; {Saikia}, D.J.
\newblock {Compact steep-spectrum and peaked-spectrum radio sources}.
\newblock {\em \aapr} {\bf 2021}, {\em 29},~3. 
\newblock {\url{https://doi.org/10.1007/s00159-021-00131-w}}.

\bibitem[{Wilkinson} et~al.(1994){Wilkinson}, {Polatidis}, {Readhead}, {Xu},
  and {Pearson}]{1994ApJ...432L..87W}
{Wilkinson}, P.N.; {Polatidis}, A.G.; {Readhead}, A.C.S.; {Xu}, W.; {Pearson},
  T.J.
\newblock {Two-sided Ejection in Powerful Radio Sources: The Compact Symmetric
  Objects}.
\newblock {\em \apjl} {\bf 1994}, {\em 432},~L87.
\newblock {\url{https://doi.org/10.1086/187518}}.

\bibitem[{Kiehlmann} et~al.(2024){Kiehlmann}, {Lister}, {Readhead}, {Liodakis},
  {O'Neill}, {Pearson}, {Sheldahl}, {Siemiginowska}, {Tassis}, {Taylor}, and
  {Wilkinson}]{2024ApJ...961..240K}
{Kiehlmann}, S.; {Lister}, M.L.; {Readhead}, A.C.S.; {Liodakis}, I.; {O'Neill},
  S.; {Pearson}, T.J.; {Sheldahl}, E.; {Siemiginowska}, A.; {Tassis}, K.;
  {Taylor}, G.B.;  et~al.
\newblock {Compact Symmetric Objects. I. Toward a Comprehensive Bona Fide
  Catalog}.
\newblock {\em \apj} {\bf 2024}, {\em 961},~240. 
\newblock {\url{https://doi.org/10.3847/1538-4357/ad0c56}}.

\bibitem[{Wu} et~al.(2017){Wu}, {Ghisellini}, {Hodges-Kluck}, {Gallo},
  {Ciardi}, {Haardt}, {Sbarrato}, and {Tavecchio}]{2017MNRAS.468..109W}
{Wu}, J.; {Ghisellini}, G.; {Hodges-Kluck}, E.; {Gallo}, E.; {Ciardi}, B.;
  {Haardt}, F.; {Sbarrato}, T.; {Tavecchio}, F.
\newblock {CMB-induced radio quenching of high-redshift jetted AGNs with highly
  magnetic hotspots}.
\newblock {\em \mnras} {\bf 2017}, {\em 468},~109--121. 
\newblock {\url{https://doi.org/10.1093/mnras/stx416}}.

\bibitem[{Vignali} et~al.(2018){Vignali}, {Piconcelli}, {Perna}, {Hennawi},
  {Gilli}, {Comastri}, {Zamorani}, {Dotti}, and {Mathur}]{2018MNRAS.477..780V}
{Vignali}, C.; {Piconcelli}, E.; {Perna}, M.; {Hennawi}, J.; {Gilli}, R.;
  {Comastri}, A.; {Zamorani}, G.; {Dotti}, M.; {Mathur}, S.
\newblock {Probing black hole accretion in quasar pairs at high redshift}.
\newblock {\em \mnras} {\bf 2018}, {\em 477},~780--790. 
\newblock {\url{https://doi.org/10.1093/mnras/sty682}}.

\bibitem[{Fu} et~al.(2015){Fu}, {Myers}, {Djorgovski}, {Yan}, {Wrobel}, and
  {Stockton}]{2015ApJ...799...72F}
{Fu}, H.; {Myers}, A.D.; {Djorgovski}, S.G.; {Yan}, L.; {Wrobel}, J.M.;
  {Stockton}, A.
\newblock {Radio-selected Binary Active Galactic Nuclei from the Very Large
  Array Stripe 82 Survey}.
\newblock {\em \apj} {\bf 2015}, {\em 799},~72. 
\newblock {\url{https://doi.org/10.1088/0004-637X/799/1/72}}.

\bibitem[{Xu} et~al.(2024){Xu}, {Cui}, {Liu}, {An}, {Cao}, {Jiang}, {Ho},
  {Chang}, {Yang}, {Shen}, {Tan}, {Han}, {Fan}, and
  {Zhang}]{2024ApJ...969...36X}
{Xu}, W.; {Cui}, L.; {Liu}, X.; {An}, T.; {Cao}, H.; {Jiang}, P.; {Ho}, L.C.;
  {Chang}, N.; {Yang}, X.; {Shen}, Y.;  et~al.
\newblock {Very Long Baseline Array Observations of Parsec-scale Radio Emission
  in Dual Active Galactic Nuclei}.
\newblock {\em \apj} {\bf 2024}, {\em 969},~36. 
\newblock {\url{https://doi.org/10.3847/1538-4357/ad463b}}.

\bibitem[{Yuan} et~al.(2018){Yuan}, {Wang}, {Worrall}, {Zhang}, and
  {Mao}]{2018ApJS..239...33Y}
{Yuan}, Z.; {Wang}, J.; {Worrall}, D.M.; {Zhang}, B.B.; {Mao}, J.
\newblock {Determining the Core Radio Luminosity Function of Radio AGNs via
  Copula}.
\newblock {\em \apjs} {\bf 2018}, {\em 239},~33. 
\newblock {\url{https://doi.org/10.3847/1538-4365/aaed3b}}.

\bibitem[{Sbarrato} et~al.(2015){Sbarrato}, {Ghisellini}, {Tagliaferri},
  {Foschini}, {Nardini}, {Tavecchio}, and {Gehrels}]{2015MNRAS.446.2483S}
{Sbarrato}, T.; {Ghisellini}, G.; {Tagliaferri}, G.; {Foschini}, L.; {Nardini},
  M.; {Tavecchio}, F.; {Gehrels}, N.
\newblock {Blazar candidates beyond redshift 4 observed by Swift}.
\newblock {\em \mnras} {\bf 2015}, {\em 446},~2483--2489. 
\newblock {\url{https://doi.org/10.1093/mnras/stu2269}}.

\bibitem[{Sotnikova} et~al.(2021){Sotnikova}, {Mikhailov}, {Mufakharov},
  {Mingaliev}, {Bursov}, {Semenova}, {Stolyarov}, {Udovitskiy}, {Kudryashova},
  and {Erkenov}]{2021MNRAS.508.2798S}
{Sotnikova}, Y.; {Mikhailov}, A.; {Mufakharov}, T.; {Mingaliev}, M.; {Bursov},
  N.; {Semenova}, T.; {Stolyarov}, V.; {Udovitskiy}, R.; {Kudryashova}, A.;
  {Erkenov}, A.
\newblock {High-redshift quasars at z {\ensuremath{\geq}} 3 -- I. Radio
  spectra}.
\newblock {\em \mnras} {\bf 2021}, {\em 508},~2798--2814. 
\newblock {\url{https://doi.org/10.1093/mnras/stab2114}}.

\bibitem[{Frey} et~al.(2024){Frey}, {Fogasy}, {Perger}, {Kulish}, {Benke},
  {Koller}, and {Gab{\'a}nyi}]{2024Univ...10...97F}
{Frey}, S.; {Fogasy}, J.; {Perger}, K.; {Kulish}, K.; {Benke}, P.; {Koller},
  D.; {Gab{\'a}nyi}, K.{\'E}.
\newblock {Revisiting a Core{\textendash}Jet Laboratory at High Redshift:
  Analysis of the Radio Jet in the Quasar PKS 2215+020 at z = 3.572}.
\newblock {\em Universe} {\bf 2024}, {\em 10},~97. 
\newblock {\url{https://doi.org/10.3390/universe10020097}}.

\bibitem[{Guo} et~al.(2025){Guo}, {An}, {Liu}, {Liu}, {Xu}, {Sotnikova},
  {Mufakharov}, and {Wang}]{2025Univ...11...91G}
{Guo}, S.; {An}, T.; {Liu}, Y.; {Liu}, C.; {Xu}, Z.; {Sotnikova}, Y.;
  {Mufakharov}, T.; {Wang}, A.
\newblock {High-Redshift Quasars at z {\ensuremath{\geq}} 3{\textemdash}III:
  Parsec-Scale Jet Properties from Very Long Baseline Interferometry
  Observations}.
\newblock {\em Universe} {\bf 2025}, {\em 11},~91. 
\newblock {\url{https://doi.org/10.3390/universe11030091}}.

\bibitem[{Koller} and {Frey}(2025)]{2025Univ...11..157K}
{Koller}, D.; {Frey}, S.
\newblock {Superluminal Motion and Jet Parameters in the High-Redshift Blazar
  J1429+5406}.
\newblock {\em Universe} {\bf 2025}, {\em 11},~157. 
\newblock {\url{https://doi.org/10.3390/universe11050157}}.

\bibitem[{Best} et~al.(1998){Best}, {Rottgering}, {Bremer}, {Cimatti}, {Mack},
  {Miley}, {Pentericci}, {Tilanus}, and {van der Werf}]{1998MNRAS.301L..15B}
{Best}, P.N.; {Rottgering}, H.J.A.; {Bremer}, M.N.; {Cimatti}, A.; {Mack},
  K.H.; {Miley}, G.K.; {Pentericci}, L.; {Tilanus}, R.P.J.; {van der Werf},
  P.P.
\newblock {Dust in 3C 324}.
\newblock {\em \mnras} {\bf 1998}, {\em 301},~L15--L19. 
\newblock {\url{https://doi.org/10.1046/j.1365-8711.1998.02155.x}}.

\bibitem[{Saxena} et~al.(2018){Saxena}, {Marinello}, {Overzier}, {Best},
  {R{\"o}ttgering}, {Duncan}, {Prandoni}, {Pentericci}, {Magliocchetti},
  {Paris}, {Cusano}, {Marchi}, {Intema}, and {Miley}]{2018MNRAS.480.2733S}
{Saxena}, A.; {Marinello}, M.; {Overzier}, R.A.; {Best}, P.N.;
  {R{\"o}ttgering}, H.J.A.; {Duncan}, K.J.; {Prandoni}, I.; {Pentericci}, L.;
  {Magliocchetti}, M.; {Paris}, D.;  et~al.
\newblock {Discovery of a radio galaxy at z = 5.72}.
\newblock {\em \mnras} {\bf 2018}, {\em 480},~2733--2742. 
\newblock {\url{https://doi.org/10.1093/mnras/sty1996}}.

\bibitem[{Drouart} et~al.(2020){Drouart}, {Seymour}, {Galvin}, {Afonso},
  {Callingham}, {De Breuck}, {Johnston-Hollitt}, {Kapi{\'n}ska}, {Lehnert}, and
  {Vernet}]{2020PASA...37...26D}
{Drouart}, G.; {Seymour}, N.; {Galvin}, T.J.; {Afonso}, J.; {Callingham}, J.R.;
  {De Breuck}, C.; {Johnston-Hollitt}, M.; {Kapi{\'n}ska}, A.D.; {Lehnert},
  M.D.; {Vernet}, J.
\newblock {The GLEAMing of the first supermassive black holes}.
\newblock {\em \pasa} {\bf 2020}, {\em 37},~e026. 
\newblock {\url{https://doi.org/10.1017/pasa.2020.6}}.

\bibitem[{Saxena} et~al.(2019){Saxena}, {R{\"o}ttgering}, {Duncan}, {Hill},
  {Best}, {Indahl}, {Marinello}, {Overzier}, {Pentericci}, {Prandoni},
  {Dannerbauer}, and {Barrena}]{2019MNRAS.489.5053S}
{Saxena}, A.; {R{\"o}ttgering}, H.J.A.; {Duncan}, K.J.; {Hill}, G.J.; {Best},
  P.N.; {Indahl}, B.L.; {Marinello}, M.; {Overzier}, R.A.; {Pentericci}, L.;
  {Prandoni}, I.;  et~al.
\newblock {The nature of faint radio galaxies at high redshifts}.
\newblock {\em \mnras} {\bf 2019}, {\em 489},~5053--5075. 
\newblock {\url{https://doi.org/10.1093/mnras/stz2516}}.

\bibitem[{Yamashita} et~al.(2020){Yamashita}, {Nagao}, {Ikeda}, {Toba},
  {Kajisawa}, {Ono}, {Tanaka}, {Akiyama}, {Harikane}, {Ichikawa}, {Kawaguchi},
  {Kawamuro}, {Kohno}, {Lee}, {Lee}, {Matsuoka}, {Niida}, {Ogura}, {Onoue}, and
  {Uchiyama}]{2020AJ....160...60Y}
{Yamashita}, T.; {Nagao}, T.; {Ikeda}, H.; {Toba}, Y.; {Kajisawa}, M.; {Ono},
  Y.; {Tanaka}, M.; {Akiyama}, M.; {Harikane}, Y.; {Ichikawa}, K.;  et~al.
\newblock {A Wide and Deep Exploration of Radio Galaxies with Subaru HSC
  (WERGS). III. Discovery of a z = 4.72 Radio Galaxy with the Lyman Break
  Technique}.
\newblock {\em \aj} {\bf 2020}, {\em 160},~60. 
\newblock {\url{https://doi.org/10.3847/1538-3881/ab98fe}}.

\bibitem[{Capetti} et~al.(2025){Capetti}, {Balmaverde}, {Coloma Puga},
  {Vizzone}, {Jimenez-Gallardo}, {Garc{\'\i}a-P{\'e}rez}, and
  {Venturi}]{2025A&A...697A.238C}
{Capetti}, A.; {Balmaverde}, B.; {Coloma Puga}, M.; {Vizzone}, B.;
  {Jimenez-Gallardo}, A.; {Garc{\'\i}a-P{\'e}rez}, A.; {Venturi}, G.
\newblock {The quest for high-redshift radio galaxies: I. A pilot spectroscopic
  study}.
\newblock {\em \aap} {\bf 2025}, {\em 697},~A238. 
\newblock {\url{https://doi.org/10.1051/0004-6361/202553967}}.

\bibitem[{Ivezi{\'c}} et~al.(2014){Ivezi{\'c}}, {Brandt}, {Fan}, {MacLeod},
  {Richards}, and {Yoachim}]{2014IAUS..304...11I}
{Ivezi{\'c}}, {\v{Z}}.; {Brandt}, W.N.; {Fan}, X.; {MacLeod}, C.L.; {Richards},
  G.T.; {Yoachim}, P.
\newblock {Optical selection of quasars: SDSS and LSST}.
\newblock In Proceedings of the Multiwavelength AGN Surveys and Studies, Yerevan, Armenia, 14--18 October 2013; 
  {Mickaelian}, A.M., {Sanders}, D.B., Eds.; IAU
  Symposium; Cambridge University Press: Cambridge, UK, 
 2014; Volume 304, pp. 11--17. 
\newblock {\url{https://doi.org/10.1017/S1743921314003159}}.

\bibitem[{Ivezi{\'c}} et~al.(2019){Ivezi{\'c}}, {Kahn}, {Tyson}, {Abel},
  {Acosta}, {Allsman}, {Alonso}, {AlSayyad}, {Anderson}, {Andrew}, {Angel},
  {Angeli}, {Ansari}, {Antilogus}, {Araujo}, {Armstrong}, {Arndt}, {Astier},
  {Aubourg}, {Auza}, {Axelrod}, {Bard}, {Barr}, {Barrau}, {Bartlett}, {Bauer},
  {Bauman}, {Baumont}, {Bechtol}, {Bechtol}, {Becker}, {Becla}, {Beldica},
  {Bellavia}, {Bianco}, {Biswas}, {Blanc}, {Blazek}, {Blandford}, {Bloom},
  {Bogart}, {Bond}, {Booth}, {Borgland}, {Borne}, {Bosch}, {Boutigny},
  {Brackett}, {Bradshaw}, {Brandt}, {Brown}, {Bullock}, {Burchat}, {Burke},
  {Cagnoli}, {Calabrese}, {Callahan}, {Callen}, {Carlin}, {Carlson},
  {Chandrasekharan}, {Charles-Emerson}, {Chesley}, {Cheu}, {Chiang}, {Chiang},
  {Chirino}, {Chow}, {Ciardi}, {Claver}, {Cohen-Tanugi}, {Cockrum}, {Coles},
  {Connolly}, {Cook}, {Cooray}, {Covey}, {Cribbs}, {Cui}, {Cutri}, {Daly},
  {Daniel}, {Daruich}, {Daubard}, {Daues}, {Dawson}, {Delgado}, {Dellapenna},
  {de Peyster}, {de Val-Borro}, {Digel}, {Doherty}, {Dubois},
  {Dubois-Felsmann}, {Durech}, {Economou}, {Eifler}, {Eracleous}, {Emmons},
  {Fausti Neto}, {Ferguson}, {Figueroa}, {Fisher-Levine}, {Focke}, {Foss},
  {Frank}, {Freemon}, {Gangler}, {Gawiser}, {Geary}, {Gee}, {Geha}, {Gessner},
  {Gibson}, {Gilmore}, {Glanzman}, {Glick}, {Goldina}, {Goldstein}, {Goodenow},
  {Graham}, {Gressler}, {Gris}, {Guy}, {Guyonnet}, {Haller}, {Harris},
  {Hascall}, {Haupt}, {Hernandez}, {Herrmann}, {Hileman}, {Hoblitt}, {Hodgson},
  {Hogan}, {Howard}, {Huang}, {Huffer}, {Ingraham}, {Innes}, {Jacoby}, {Jain},
  {Jammes}, {Jee}, {Jenness}, {Jernigan}, {Jevremovi{\'c}}, {Johns}, {Johnson},
  {Johnson}, {Jones}, {Juramy-Gilles}, {Juri{\'c}}, {Kalirai}, {Kallivayalil},
  {Kalmbach}, {Kantor}, {Karst}, {Kasliwal}, {Kelly}, {Kessler}, {Kinnison},
  {Kirkby}, {Knox}, {Kotov}, {Krabbendam}, {Krughoff}, {Kub{\'a}nek},
  {Kuczewski}, {Kulkarni}, {Ku}, {Kurita}, {Lage}, {Lambert}, {Lange},
  {Langton}, {Le Guillou}, {Levine}, {Liang}, {Lim}, {Lintott}, {Long},
  {Lopez}, {Lotz}, {Lupton}, {Lust}, {MacArthur}, {Mahabal}, {Mandelbaum},
  {Markiewicz}, {Marsh}, {Marshall}, {Marshall}, {May}, {McKercher}, {McQueen},
  {Meyers}, {Migliore}, {Miller}, and {Mills}]{2019ApJ...873..111I}
{Ivezi{\'c}}, {\v{Z}}.; {Kahn}, S.M.; {Tyson}, J.A.; {Abel}, B.; {Acosta}, E.;
  {Allsman}, R.; {Alonso}, D.; {AlSayyad}, Y.; {Anderson}, S.F.; {Andrew}, J.;
  et~al.
\newblock {LSST: From Science Drivers to Reference Design and Anticipated Data
  Products}.
\newblock {\em \apj} {\bf 2019}, {\em 873},~111. 
\newblock {\url{https://doi.org/10.3847/1538-4357/ab042c}}.

\end{thebibliography}
\end{document}